\documentclass[twocolumn]{aastex701}

\usepackage{CJKutf8}

\graphicspath{{./}{figures/}}

\usepackage{hyperref}

\shortauthors{liu et al.}
\newcommand{\emailaddress}{liuyuhua@shao.ac.cn}

\shorttitle{ALMA Polarization Observation of the 20\,\kms{} Cloud}

\newcommand{\dotarcsec}{\rlap{.}\arcsec}
\usepackage{csquotes}
\usepackage{amsmath,amsthm,amssymb}
\usepackage{mathtools}
\usepackage[inline]{enumitem} 
\usepackage{float}
\graphicspath{{./}{figures/}}
\usepackage{graphicx}
\usepackage{booktabs}
\usepackage{nameref}
\usepackage{rotating}

\usepackage{mathrsfs}

\usepackage[nameinlink]{cleveref}
\Crefname{figure}{Figures}{Figures}

\newcommand{\kms}{\mbox{km\,s$^{-1}$}}
\usepackage[title]{appendix}

\begin{document}
\begin{CJK}{UTF8}{gbsn}

\title{ALMA Polarization Study of the Magnetic Fields in Two Massive Clumps in the 20\,km\,s$^{-1}$ Cloud of the Central Molecular Zone}

\correspondingauthor{Yuhua Liu}
\email{\emailaddress}

\author[0009-0009-2263-5502]{Yuhua Liu (柳玉华)}
\affiliation{Shanghai Astronomical Observatory, Chinese Academy of Sciences, 80 Nandan Road, Shanghai 200030, P.\ R.\ China}
\email{liuyuhua@shao.ac.cn}

\author[0000-0003-2619-9305]{Xing Lu (吕行)}
\affiliation{Shanghai Astronomical Observatory, Chinese Academy of Sciences, 80 Nandan Road, Shanghai 200030, P.\ R.\ China}
\affiliation{State Key Laboratory of Radio Astronomy and Technology, A20 Datun Road, Chaoyang District, Beijing, 100101, P.\ R.\ China}
\email{xinglu@shao.ac.cn}

\author[0000-0002-4774-2998]{Junhao Liu (刘峻豪)}
\affiliation{National Astronomical Observatory of Japan, 2-21-1 Osawa, Mitaka, Tokyo, 181-8588, Japan}
\email{liujunhao42@outlook.com}

\author[0000-0003-1337-9059]{Xing Pan (潘兴)}
\affiliation{School of Astronomy and Space Science, Nanjing University, 163 Xianlin Avenue, Nanjing 210023, P.\ R.\ China}
\affiliation{Key Laboratory of Modern Astronomy and Astrophysics (Nanjing University), Ministry of Education, Nanjing 210023, P.R.China}
\affiliation{Center for Astrophysics $\vert$ Harvard \& Smithsonian, 60 Garden Street, Cambridge, MA, 02138, USA}
\email{xingpan1017@gmail.com}

\author[0000-0003-2384-6589]{Qizhou Zhang}
\affiliation{Center for Astrophysics $\vert$ Harvard \& Smithsonian, 60 Garden Street, Cambridge, MA, 02138, USA}
\email{qzhang@cfa.harvard.edu}

\author[0000-0003-2300-2626]{Hauyu Baobab Liu}
\affiliation{Department of Physics, National Sun Yat-Sen University, No. 70, Lien-Hai Road, Kaohsiung City 80424}
\affiliation{Center of Astronomy and Gravitation, National Taiwan Normal University, Taipei 116}
\email{baobabyoo@gmail.com}

\author[0009-0003-5699-2723]{Meng-Zhe Yang}
\affiliation{Institute of Astronomy and Department of Physics, National Tsing Hua University, Hsinchu 30013, Taiwan}
\email{jeremys93102@gapp.nthu.edu.tw}

\author[0000-0001-5522-486X]{Shih-Ping Lai}
\affiliation{Institute of Astronomy and Department of Physics, National Tsing Hua University, Hsinchu 30013, Taiwan}
\email{slai@gapp.nthu.edu.tw}

\author[0000-0001-8516-2532]{Tao-Chung Ching}
\affiliation{National Radio Astronomy Observatory, P.O. Box O, Socorro, NM 87801, USA}
\email{chingtaochung@gmail.com}

\author[0000-0001-9822-7817]{Wenyu Jiao (焦文裕)}
\affiliation{Shanghai Astronomical Observatory, Chinese Academy of Sciences, 80 Nandan Road, Shanghai 200030, P.\ R.\ China}
\email{wenyujiao@shao.ac.cn}

\author[0000-0001-7817-1975]{Yankun Zhang (张燕坤)}
\affiliation{Shanghai Astronomical Observatory, Chinese Academy of Sciences, 80 Nandan Road, Shanghai 200030, P.\ R.\ China}
\email{zhangyankun@shao.ac.cn}

\author[0000-0001-8077-7095]{Pak Shing Li}
\affiliation{Shanghai Astronomical Observatory, Chinese Academy of Sciences, 80 Nandan Road, Shanghai 200030, P.\ R.\ China}
\email{pakshingli@shao.ac.cn}

\author[0000-0003-3540-8746]{Zhiqiang Shen (沈志强)}
\affiliation{Shanghai Astronomical Observatory, Chinese Academy of Sciences, 80 Nandan Road, Shanghai 200030, P.\ R.\ China}
\affiliation{State Key Laboratory of Radio Astronomy and Technology, A20 Datun Road, Chaoyang District, Beijing, 100101, P.\ R.\ China}
\email{zshen@shao.ac.cn}

\author[0000-0002-5286-2564]{Tie Liu (刘铁)}
\affiliation{Shanghai Astronomical Observatory, Chinese Academy of Sciences, 80 Nandan Road, Shanghai 200030, P.\ R.\ China}
\affiliation{State Key Laboratory of Radio Astronomy and Technology, A20 Datun Road, Chaoyang District, Beijing, 100101, P.\ R.\ China}
\email{liutie@shao.ac.cn}

\author[0000-0001-7817-1975]{Adam Ginsburg}
\affiliation{University of Florida Department of Astronomy, Bryant Space Science Center, Gainesville, FL, 32611, USA}
\email{adamginsburg@ufl.edu}

\author[0000-0002-2826-1902]{Qi-lao Gu (顾琦烙)}
\affiliation{Shanghai Astronomical Observatory, Chinese Academy of Sciences, 80 Nandan Road, Shanghai 200030, P.\ R.\ China}
\email{qlgu@shao.ac.cn}

\author[0000-0003-0596-6608]{Mengke Zhao}
\affiliation{School of Astronomy and Space Science, Nanjing University, 163 Xianlin Avenue, Nanjing 210023, P.\ R.\ China}
\email{mkzhao628@gmail.com}
\affiliation{Key Laboratory of Modern Astronomy and Astrophysics (Nanjing University), Ministry of Education, Nanjing 210023, P.R.China}

\begin{abstract} 
We present the Atacama Large Millimeter/submillimeter Array (ALMA) observations of linearly polarized 870\,$\mu$m continuum emission at a resolution of $\sim$0\dotarcsec2 (2000 au) toward the two massive clumps, Clump\,1 and Clump\,4, in the 20\,km\,s$^{-1}$ cloud. The derived magnetic field strengths for both clumps range from $\sim$0.3 to 3.1\,mG using the Angular Dispersion Function (ADF) method. The magnetic field orientations across multiple scales suggest that the magnetic field dominates at the cloud scale, whereas gravity likely governs structures at the core (0.01$-$0.1 pc) and condensation ($\le$ 0.01 pc) scales. Furthermore, the study on the angular difference between the orientations of the local gravity gradient and the magnetic field suggests that the gravity predominantly governs the dynamics in the diffuse regions, while both gravity and star formation feedback become increasingly significant within the dense regions. The ratio of the magnetic field tension force $F_\textnormal{B}$ to the gravitational force $F_\textnormal{G}$ suggests that the magnetic field may provide some support against gravity, but it is insufficient to prevent gas from infalling toward the dense cores.
\end{abstract}


\section{Introduction}

The central molecular zone \citep[CMZ;][]{morris1996} is the innermost region of the Milky Way with a Galactocentric radius of $\sim$300\,pc \citep{henshaw2023}. This region contains a large amount of molecular gas of $\gtrsim$10$^{7}$\,M$_{\odot}$ \citep[e.g.,][]{dahmen1998, longmore2013a,battersby2024}, and represents a star-forming environment under some extreme physical conditions. The magnitudes of the average number density of molecular gas \citep[$\gtrsim$10$^{4}$\,cm$^{-3}$; e.g.,][]{bally1987,ferri2007,longmore2013a}, gas temperatures \citep[$\sim$25 K to $>$100\,K; e.g.,][]{ao2013,ginsburg2016,immer2016,krieger2017}, Mach numbers \citep[10--20; e.g.,][]{federrath2016,henshaw2016}, and magnetic field strengths \citep[$\sim$0.1~mG to $\gtrsim$ 1~mG; e.g.,][]{pillai2015,mangilli2019,lu2024,pan2024, pan2025, pare2025} in this region are found to be higher than those in the Galactic disk. However, the star formation rate (SFR) is about one order of magnitude lower than expected, considering the mass and density of the dense gas in this region \citep[e.g.,][]{longmore2013a,barnes2017}.

Observations have identified several massive molecular clouds within the CMZ, such as G0.253+0.016, the 20\,km\,s$^{-1}$ cloud, the 50\,km\,s$^{-1}$ cloud, Sgr~C, and Sgr~B2 \citep[e.g.,][]{kauffmann2017}. Some of these clouds except for Sgr B2 have shown lower SFRs than predicted by the dense gas star formation relation derived from the solar neighborhood \citep[e.g.,][]{kauffmann2017, lu2019ApJS}. \cite{kruijssen2014} suggested that the low star SFR in the CMZ is primarily due to the strong turbulence. Later, \cite{lu2024} emphasized that turbulence and self-gravity play the important roles in star formation based on the observed weak correlation between magnetic field strength and SFR. More recently, \cite{yang2025} indicated that weak correlation between magnetic field strength and SFR could imply that the SFR is not linearly correlated with magnetic field
strength and suggested that the dominance of magnetic fields may also play a crucial role in suppressing the star formation in this region. Furthermore, \cite{pan2025} presented a SOFIA data for the CMZ revealed a parallel configuration of the relative orientation between the magnetic field and column density and a supervirial state for all CMZ clouds except for Sgr\,B2, suggesting that self-gravity may not be dominant even in the most dense structures. The role of the magnetic field and in particular its interplay with other sources such as gravity in star formation remains unclear. All above studies were conducted at spatial resolutions of 0.2$-$0.5\,pc. To address this degeneracy, we present high-resolution polarization observations that trace magnetic field morphology at an unprecedented resolution of $\sim$0.01\,pc (2000 au).

In this work, we study the magnetic field properties and morphologies through the Atacama Large Millimeter/submillimeter Array (ALMA) high angular resolution polarization observations toward two massive clumps, named as Clump 1 and Clump 4, in the 20\,km\,s$^{-1}$ cloud (see \autoref{fig1-cont-B-vec}\,(a)), one of the massive molecular clouds \citep[$\gtrsim10^5 M_{\odot}$;][]{lu2015,lu2017} in the CMZ, which is highly turbulent and also relatively more active in star formation compared to the other clouds \citep{kauffmann2017,lu2019}. We aim to characterize the magnetic field properties toward this cloud.

\section{Observations and Data Reduction} \label{sec:obser}

\subsection{ALMA observations}
\begin{figure*}[h]
\includegraphics[width=0.98\textwidth]{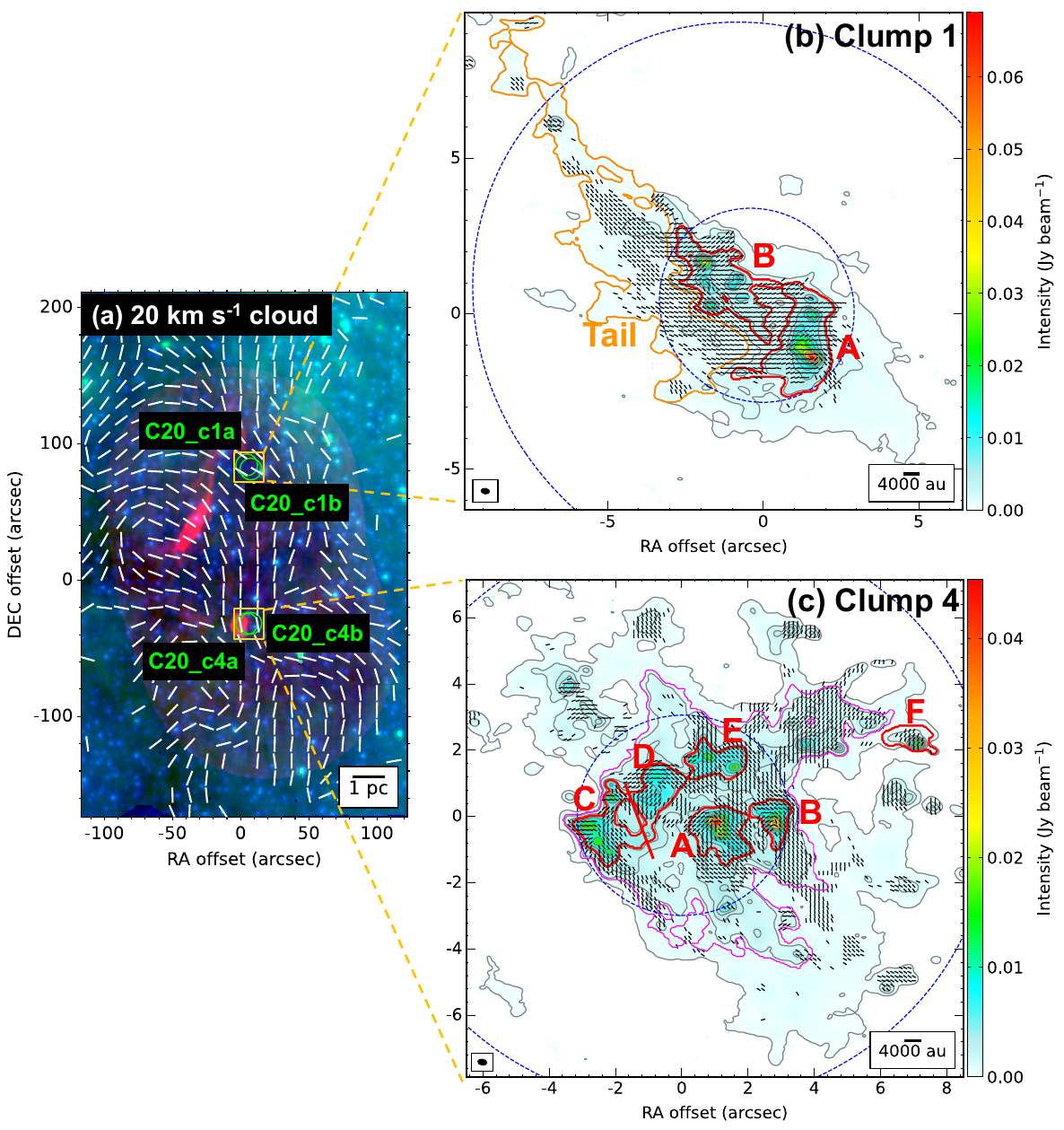}
\caption{Left: false-color image of the 20\,km\,s$^{-1}$ cloud created using the MeerKAT 1.28~GHz continuum in red \citep{heywood2022}, Spitzer 3.6\,$\mu$m in blue, and Spitzer 8\,$\mu$m in green \citep{stolovy2006}. The white line segments are the inferred magnetic field orientation obtained from James Clerk Maxwell Telescope (JCMT) observations of polarized dust emission \citep{yang2025}. The ALMA observation fields are denoted by the lime circles. Right: The inferred magnetic field orientation (black line segments) overlaid on Stokes $I$ emission (color) toward (b) Clump\,1 and (c) Clump\,4. The grey contours are the Stokes $I$ emission, which correspond to [5, 25, 50, 100, 150, 200, 350, 500, 650] $\times$ $\sigma$ (1$\sigma$ = 65~$\mu$Jy\,beam$^{-1}$) in panel (b) and [5, 25, 50, 75, 100, 150, 250, 350, 450, 650] $\times$ $\sigma$ (1$\sigma$ = 48~$\mu$Jy\,beam$^{-1}$) in panel (c). The magenta contour in panel (c) denote the area used to derive the magnetic field properties for Clump\,4. The cores enclosed by red solid contours in these two clumps are identified by \texttt{astrodendro}. Clump\,1-tail is enclosed using the orange solid contour, which is identified as the diffuse emission extended along the northeast-southwest direction that correspond to $\sim$5$\sigma$ in panel (b). Clump\,4-C and Clump\,4-D, identified as one core using \texttt{astrodendro}, are manually divided by the red solid line using the 75$\sigma$ contour as the reference. The synthesized beam size is denoted by filled black ellipse in the bottom left corner. The vectors in all panels are selected using Super-Nyquist sampling \cite[$\sim$3 pixels per beam;][]{hull2020} and set to have a uniform length. The outer and inner blue dashed contours in panels (b) and (c) are the FHWM of the primary beam and the inner 1/3 of it, respectively. The absolute coordinates of the origins are (a) [266.408$^{\circ}$, $-$29.0858$^{\circ}$], (b) [266.407$^{\circ}$, $-$29.0635$^{\circ}$], and (c) [266.407$^{\circ}$, $-$29.0956$^{\circ}$], respectively. } 
\label{fig1-cont-B-vec}
\end{figure*}

The ALMA Band 7 (870\,$\mu$m) full polarization observations targeting seven fields in the CMZ, among which four fields are inside the 20\,km\,s$^{-1}$ cloud (see the lime circles in \autoref{fig1-cont-B-vec}(a)), were obtained in Cycle 8 (Project ID: 2021.1.00286.S; PI: X.\ Lu) using two array configurations. The higher resolution observations ($\sim$0\dotarcsec24$\times$0\dotarcsec16) were obtained on 2022 June 15 and 16 in two consecutive observing blocks, using the more extended C-5 configuration with a projected baseline range of 15\,m to 1.3\,km. The lower resolution observations ($\sim$0\dotarcsec71$\times$0\dotarcsec54) were obtained on 2022 April 10 in one observing block, using the compact C-2 configuration with a projected baseline range of 15 to 455\,m. The on-source time for each field in the C-5 and C-2 configurations is 114 and 30 minutes, receptively. Between 44 and 48 of the 12-m antennas were operated during the observations. The full width at half maximum (FWHM) of the ALMA primary beam is $\sim$17$\arcsec$. The maximum recoverable size is $\sim$6.1$\arcsec$. Four spectral windows (SPWs) centered at 334.2, 336.1, 346.2, and 348.1 GHz, each with a total bandwidth of 1.875 GHz, were allocated with the frequency division mode for the continuum observations. The raw data were first calibrated with the standard ALMA pipeline using the Common Astronomy Software Applications \citep[CASA;][]{bean2022casa} version 6.2.1-7, followed by polarization calibrations done by a customized calibration script provided by the ALMA staff.

Furthermore, two additional executions in the C-2 configuration were obtained on 2022 March 31, with the baselines ranging from 14 to 313\,m. The data are labeled as Quality Assurance (QA) semipass due to a phase loop lock (PLL) error of ALMA. As described in X.\ Lu et al. (2026, in preparation), the outcomes of the QA-pass and QA-semipass data are consistent, hence the QA-semipass data were incorporated in the final imaging to improve the continuum sensitivity.

We began by inspecting the calibrated data and flagging channels with line contamination. Then, the flagged data from the two array configurations, including the QA-semipass data for each field C20$\_$c1a, C20$\_$c1b, C20$\_$c4a, and C20$\_$c4b (see \autoref{fig1-cont-B-vec}\,(a)) were combined using the CASA task \texttt{concat}. In order to enhance the image quality for Stokes $I$, we performed three rounds of phase self-calibration. We derived the self-calibration solutions to each field separately. 
The first round of self-calibration was derived with a solution interval equivalent to the scan length ($\sim$60~s). The second and third rounds were derived with the solution intervals equal to the integration time ($\sim$6~s). Then, we combined the fields C20$\_$c1a and C20$\_$c1b, as well as C20$\_$c4a and C20$\_$c4b, to produce mosaicked continuum images for Clump\,1 and Clump\,4, respectively, as shown in \Cref{fig1-cont-B-vec}\,(b) and (c). The Stokes $I$ images were made using the data with self-calibration applied, whereas the Stokes $Q$ and $U$ images were made using the data without self-calibration. The final CLEANed images of Stokes $I$, $Q$, and $U$ for each clump were made using the CASA task \texttt{tclean} with Briggs weighting (robust = 0.5) and multi-scale algorithm with scales of [0, 5, 15, 50, 150] pixels, each pixel having a size of 0\dotarcsec04.
The debiased linearly polarized intensity (PI) was derived from Stokes $Q$ and $U$ as $\textnormal{PI}=\sqrt{Q^2+U^2-\sigma_{QU}^{2}}$, where $\sigma_{QU}=\sqrt{\sigma_{Q}^2+\sigma_{U}^2}$. The position angle (P.A.) of the polarization vector \(\chi=\frac{1}{2}\textnormal{arctan}(U/Q)\) was computed using the CASA task \texttt{immath}. Each of the Stokes $I$ and polarization vector images was obtained using pixels where an emission detection is $\geqslant$ 3$\sigma_{QU}$.

In order to compare the magnetic field morphology at different scales, we additionally imaged the continuum emission for both Clump\,1 and Clump\,4 using the extended C-5 and compact C-2 configurations separately. These images were made using the self-calibrated data as described above. \autoref{tab:img-statistics} summarizes the rms noise level measured in each Stokes component, the synthesized beam, and P.A.\ for each of the continuum images.

In addition to the continuum images, we also imaged each SPW in these datasets to search for possible lines that could trace the turbulence. The HN$^{13}$C ($J$ = 4$-$3) line image will be used to derive the velocity dispersion. The details can be found in \autoref{apdx-B-field}.

\begin{table}[ht!]
\scriptsize 
\caption{Summary of the images}
\label{tab:img-statistics}
\centering
\begin{tabular}{lcccc}
\hline\hline
Image & rms Noise Level & Synthesized & P.A. &Figure\\
Name& Stokes $I$, $Q$, $U$ & Beam Size &  &\\
&($\mu$Jy\,beam$^{-1}$)&(\arcsec $\times$ \arcsec)& ($^\circ$)& \\
\hline
Clump\,1 & 65, 18, 18&  0.31$\times$0.21 &81& \ref{fig1-cont-B-vec}\,(b)\\
Clump\,4 & 48, 17, 17& 0.31$\times$0.21 &81 &\ref{fig1-cont-B-vec}\,(c)\\
\hline

Clump\,1(C-2)& 220, 33, 33 & 0.71$\times$0.54  &81  & \ref{jcmt-alma-c1}\,(b)\\

Clump\,1(C-5)& 32, 21, 21  & 0.24$\times$0.17  &84  & \ref{jcmt-alma-c1}\,(c)--(d)\\
\hline

Clump\,4(C-2)& 210, 33, 33  & 0.71$\times$0.55  &81  & \ref{jcmt-alma-c4}\,(b)\\

Clump\,4(C-5)& 22, 21, 21  & 0.24$\times$0.16  &84  & \ref{jcmt-alma-c4}\,(c)--(h)\\
\hline
\end{tabular}
\end{table}

\subsection{JCMT SCUBA-2/POL-2 Observations}
We utilized the data obtained from James Clerk Maxwell Telescope (JCMT) observations toward the 20\,km\,s$^{-1}$ cloud \citep{yang2025,karoly2025} to compare with the ALMA data in this work. The 850$\mu$m continuum and polarization observations were conducted using the Common-User Bolometer Array 2 (SCUBA-2) with the POL-2 polarimeter on the JCMT as part of the B-fields In STar forming Regions Observations (BISTRO) survey \citep[Project ID: M20AL018;][]{ward-Thompson2017}. The beam size is 14\dotarcsec6 \citep{dempsey2013}. The SCUBA-2/POL-2 raw data were reduced using the SubMillimetre User Reduction Facility (SMURF) package with \texttt{pol2map} routine \citep{chapin2013}. The details of the data reduction were described by \cite{karoly2025}. We obtained the Stokes $I, Q, U$ images from \cite{yang2025} and the debiased polarization catalog was binned to 12\arcsec \citep{yang2025}.

\section{Results}\label{sec3.1}

\subsection{870\,$\mu$m Dust Continuum (Stokes I)}
\Cref{fig1-cont-B-vec}\,(b) and (c) show 870\,mm continuum emissions for Clump\,1 and Clump\,4 overlaid with the inferred magnetic field orientations (by rotating the polarization orientation by 90$^{\circ}$) assuming that the dust alignment is due to the magnetic field \citep[e.g.,][]{hildebrand1988, andersson2015}. These two clumps are split into substructures (cores) using the \texttt{astrodendro} Python package \citep{rosolowsky2008}. Here, a ``core'' refers to the branch identified by \texttt{astrodendro}, and a ``condensation'' refers to the isolated compact structure within the core. The details for the dendrogram parameters can be found in \autoref{apdx-astro}.

We first estimated the physical properties. The gas mass $M_{\textnormal{gas}}$ for each core was estimated as
\begin{equation}
    M_{\textnormal{gas}}=\Lambda\frac{F_{\nu}d^{2}}{\kappa_{\nu}B_{\nu}(T)},
\end{equation}
where $F_{\nu}$ is the total flux density (with primary-beam-corrected) provided by \texttt{astrodendro} for all cores (for Clump\,1-tail, $F_{\nu}$ is manually measured from the Stokes $I$ image using the CASA task \texttt{imstat} with the key \texttt{[flux]}), $d=8.277$\,kpc is adopted as the distance to the CMZ assuming that the distance to the CMZ is the same as that to the super massive black hole \citep{GRAVITY2022}, $\Lambda=100$ is the gas-to-dust mass ratio, $\kappa_{\nu}=1.69$\,cm$^{2}$g$^{-1}$ is the dust opacity interpolated at $\lambda=870\,\mu$m from the coagulation model of dust grains coated with thin ice mantles at a gas density of 10$^{6}$\,cm$^{-3}$ \citep{ossenkopf1994}, $B_{\nu}(T)$ is the Planck function as a function of the dust temperature $T$. We adopted a uniform dust temperature of $T=20$\,K in the CMZ \citep{Pierce-Price2000,tang2021,battersby2025}. The column density $N{_{\textnormal{H}_{2}}}$ and volume density $n{_{\textnormal{H}_{2}}}$ were then estimated as
\begin{equation}
N{_{\textnormal{H}_{2}}}=\frac{M_{\textnormal{gas}}}{\pi r^{2}\mu_{\textnormal{H}_{2}}m_{\textnormal{H}}}, 
\end{equation}
and
\begin{equation}
n{_{\textnormal{H}_{2}}}=\frac{3M_{\textnormal{gas}}}{4\pi r^{3}\mu_{\textnormal{H}_{2}}m_{\textnormal{H}}}=\frac{3N_{\textnormal{H}_{2}}}{4r},
\end{equation}
respectively, where $r$ is the effective radius, $r=\sqrt{A/\pi}$, where $A$ is the exact area of the core, $\mu_{\textnormal{H}_{2}}=2.8$ is the mean molecular weight per hydrogen molecule \citep{kauffmann2008}, and $m_{\textnormal{H}}=1.67\times10^{-24}\,\textnormal{g}$ is the mass of the hydrogen atom. We assumed a spherical geometry with a volume given by $V=4/3\pi r^3$. The physical quantities $r$, $F_{\nu}$, $M_{\textnormal{gas}}$, $N{_{\textnormal{H}_{2}}}$, and $n{_{\textnormal{H}_{2}}}$ are summarized in \autoref{tab:physical}.

The gas masses of all cores and Clump\,1-tail range from 12 to 240\,M$_{\odot}$. All the cores show a typical volume density of 10$^{6}$\,cm$^{-3}$ and a typical column density of 10$^{23}$cm$^{-2}$, respectively. 

\subsection{Magnetic Field Strength}\label{sec:result-B-field}

We estimated the plane-of-sky (POS) total magnetic field strengths (B$_{\textnormal{pos}}$) toward Clump\,1 and Clump\,4 using the Angular Dispersion Function (ADF) method. The ADF method provides estimates of the turbulent-to-ordered magnetic field strength ratio based on the angular dispersion function of the magnetic field position angle, where the angle differences are evaluated as a function of the distance between pairs of points \citep{hildebrand2009,houde2009,houde2016}. By using the method developed by \cite{hildebrand2009}, $B_{\textnormal{pos}}$ given by \cite{liujunhao2021,liujunhao2022a} is
\begin{equation}
\label{eq:adf-hil09}
B_{\textnormal{pos}}\sim0.21\sqrt{4\pi\rho}\,\sigma_{v}\left( \frac{\langle B_\textnormal{t}^{2} \rangle}{\langle B^{2} \rangle}\right)^{-\frac{1}{2}}_{\textnormal{int}},
\end{equation}
where $(\langle B_\textnormal{t}^{2} \rangle/\langle B^{2} \rangle)^{-1/2}_{\textnormal{int}}$ is the turbulent-to-total field strength ratio, without correction for line-of-sight (LOS) signal integration. $B_{\textnormal{t}}$ denotes the turbulent magnetic field and $B$ denotes the total magnetic field, which the total magnetic field is the sum of the ordered magnetic field and turbulent magnetic field. 0.21 is the average correction factor for the spherical and cylindrical structures \citep{liujunhao2021}. $\rho=n_{\textnormal{H}_{2}}\mu_{\textnormal{H}_{2}}m_{\textnormal{H}}$ is the mass density. The turbulent velocity dispersion $\sigma_{\nu}$ is shown in \autoref{v-dispersion}. The polarization angle distributions $\sigma_{\theta}$ for the entire Clump\,1 and Clump\,4 are shown in \autoref{clump-angle}. The polarization angle distributions for the cores within each clump are shown in \autoref{sigmat}. The ADF is shown in \autoref{fig-adf}. The derivation of $\langle B_\textnormal{t}^{2} \rangle/\langle B^{2}\rangle$ can be found in \autoref{apdx-B-field}. The cores in Clump\,1 have sufficient independent measurements \citep[$>$28;][]{liujunhao2021}. In Clump\,4, however, only the Clump\,4-A core has sufficient independent measurements, while all other cores have fewer than 28 independent measurements. Therefore, we only estimated the magnetic field properties for Clump\,4-A. In addition, we defined an enclosed area within Clump\,4, where polarization detections are concentrated, with more than 70\% of all polarization detections above the 5$\sigma$ level originating there. This selected region was identified by \texttt{astrodendro} as a branch included most of the cores within this clump (enclosed by the magenta contour in \autoref{fig1-cont-B-vec}). The magnetic field properties derived for this enclosed area are taken to represent those of the Clump\,4 as a whole. 

In this work, we do not use the classical Davis–Chandrasekhar–Fermi (DCF) method to derive the magnetic field strength due to the large angle dispersion in our data especially in Clump\,4 as shown in \autoref{clump-angle} and \autoref{sigmat}. In addition, the ADF method accounts for the order field and observational effects \cite[e.g.,][]{hildebrand2009,houde2009,houde2016} and \cite{liujunhao2021} found that the uncertainty of the ADF method is smaller than DCF.

\begin{table*}[ht]
{\scriptsize 
\begin{center}
\caption{Physical Properties in the Two Clumps}
\label{tab:physical}
\begin{tabular}{ccccccccccc}
\hline\hline \noalign {\smallskip}
Parameter& Clump\,1-A   &Clump\,1-B &Clump\,1-tail&Clump\,4-A&Clump\,4-B&Clump\,4-C&Clump\,4-D&Clump\,4-E&Clump\,4-F& Clump\,4\\
\hline \noalign {\smallskip}
$r$ [pc]& 0.052  &0.049 & 0.108 & 0.033& 0.023& 0.024& 0.028& 0.027& 0.021 &0.146   \\
$F_{\nu}$ [mJy]& 593.4 &  369.5  &  273.6  &  251.1 &129.6 & 137.4 & 134.6 &132.2 & 29.5&1809.5   \\
$M_{\textnormal{gas}}$ [$M_{\odot}$]& 245.6  & 152.9    & 113.3  & 103.9
 & 53.6 & 56.9  &  55.7& 54.7 &  12.2 &749.0  \\
 $N{_{\textnormal{H}_{2}}}$ [$10^{23}$ cm$^{-2}$]& 13.0  &   9.1  & 1.4&  13.6&14.4 &14.0 &10.1 &10.7 &3.9 &5.0      \\
$n{_{\textnormal{H}_{2}}}$ [$10^{6}$ cm$^{-3}$]& 6.0  &  4.5   &  0.3   &10.0 & 15.2 & 14.2 & 8.8 & 9.6 & 4.6 & 0.8  \\
$\sigma_{\theta}$ [deg]$^a$ & 11.0 $\pm$ 0.5 &22.0$\pm$1.8 & 15.8$\pm$0.9  
&15.0$\pm$1.4 &19.6$\pm$2.6 
&35.5$\pm$7.7 &98.7$\pm$67.1 
&15.1$\pm$1.3 &4.6$\pm$0.1 &-\\
$\sigma_{v}$ [km\,s$^{-1}$]& 
1.68$\pm$0.64  &1.45$\pm$0.20& 2.25$\pm$0.27$^b$& 
1.16$\pm$0.17&1.38$\pm$0.08 &0.82$\pm$0.12&
1.03$\pm$0.37&1.13$\pm$0.34& 0.96$\pm$0.31&1.08$^c$\\
\hline \noalign {\smallskip}

\end{tabular}
\end{center}
}
\footnotesize $^a${The angle dispersion is derived using the regrided image (3 pixels per beam).}\\
\footnotesize $^b${The velocity dispersion for Clump\,1-tail is derived from NH$_{3}$.}\\
\footnotesize $^c${The velocity dispersion for Clump\,4 is derived by averaging the dispersions from all its cores.}\\
\end{table*}

\begin{table*}[ht!]
{\scriptsize 
\begin{center}
\caption{Magnetic Field Strengths and Properties}
\label{tab:b-field}
\begin{tabular}{cccccc}
\hline\hline \noalign {\smallskip}
Parameter& Clump\,1-A   &Clump\,1-B &Clump\,1-tail$^a$  &Clump\,4-A&Clump\,4\\
\hline \noalign {\smallskip}

$B_{\textnormal{pos}}$ [mG]&
 3.1& 2.3 & 1.8 &
2.3 &0.3\\
$\lambda$&2.6&2.4&0.5&3.7&9.0
\\
$\mathcal{M}_{\textnormal{A}}$&1.4& 1.4& 0.7 &1.7& 3.0\\
\hline \noalign {\smallskip}
$\sigma_{\textnormal{tot}}$ [km\,s$^{-1}$]& 
1.70&1.47&2.27&1.19&1.11\\
$M_{\textnormal{k+B}}$[$M_\odot$]&179.6&123.5&-&54.7&192.4\\
$\alpha_{\textnormal{k+B}}$&0.7&0.8&-&0.5&0.3\\
\hline \noalign {\smallskip}

\end{tabular}
\end{center}
}
\footnotesize $^a${$M_{\textnormal{k+B}}$ and $\alpha_{\textnormal{k+B}}$ are not calculated for Clump\,1-tail as it is an elongated diffuse structure.}\\
\end{table*}

\subsection{Mass-to-flux Ratios}\label{subsec:results_mtfr}
The relative importance of gravity and magnetic fields can be accessed using the mass-to-flux ratio normalized to its critical value, which is described by a dimensionless parameter $\lambda$ and given by \cite{crutcher2004}
\begin{equation}
\begin{split}
\lambda=\frac{(M/\Phi)}{(M/\Phi)_{\textnormal{cr}}}&=\frac{\pi r^2 \mu_{\textnormal{H}_{2}}m_{\textnormal{H}}N_{\textnormal{H}_{2}}/\pi r^2 B_\textnormal{3D}}{1/(2\pi\sqrt{G})}\\
&=\frac{\mu_{\textnormal{H}_{2}}m_{\textnormal{H}}N_{\textnormal{H}_{2}}(2\pi\sqrt{G})}{B_\textnormal{3D}},
\end{split}
\end{equation}
where $M=\pi r^2 \mu_{\textnormal{H}_{2}}m_{\textnormal{H}}N_{\textnormal{H}_{2}}$ is the observed gas mass $M_{\textnormal{gas}}$, $\Phi=\pi r^2 B_\textnormal{3D}$ is the magnetic flux, $B_\textnormal{3D}$ is the three-dimensional (3D) total magnetic field strength, $(M/\Phi)_{\textnormal{cr}}=1/(2\pi\sqrt{G})$ is the critical mass-to-flux ratio, and $G$ is the gravitational constant. Since the line-of-sight magnetic field strength is not constrained by polarization measurements, we estimated the 3D total magnetic field strength $B_\textnormal{3D}$ from $B_{\textnormal{pos}}$ by adopting a statistical relation $B_\textnormal{3D}\sim1.25\,B_{\textnormal{pos}}$ \citep{liujunhao2022b}, assuming the uncertainties of factor of 2 \citep{liujunhao2021}. The factor of 1.25 is the mean value of the correlation factors derived from the statistical relation between the 3D and POS uniform magnetic field \citep[$B_{\textnormal{u,3D}}=\frac{4}{\pi}\,B_{\textnormal{u,pos}}$;][]{crutcher2004}, and the statistical relation between the 3D and POS turbulent magnetic field \citep[$B_{\textnormal{t,3D}}=\sqrt{\frac{3}{2}}\,B_{\textnormal{t,pos}}$;][]{liujunhao2021}, respectively. The estimated $\lambda$ values are presented in \autoref{tab:b-field}. $\lambda<1$ indicates a magnetically subcritical state, where the magnetic field dominates over gravity and prevents gravitational collapse; $\lambda=1$ indicates a critical state, in which the magnetic field and gravity are balanced and the system is on the verge of stability; $\lambda>1$ denotes a magnetically supercritical state, where the magnetic field cannot overcome gravity and the system collapses. Clump\,1-A, Clump\,1-B, and Clump\,4-A have $\lambda>1$, suggesting that they are magnetically supercritical. Clump\,1-tail has $\lambda\sim0.5$, indicating that it is subcritical.

\subsection{Alvénic Mach Numbers}
The balance between the magnetic field and turbulence can be characterized by the Alvénic Mach number
\begin{equation}
\mathcal{M}_{\textnormal{A}}=\sigma_{v,3\textnormal{D}}/v_{\textnormal{A,3D}},
\end{equation}
where $\sigma_{v,3\textnormal{D}}=\sqrt{3}\sigma_{v}$ is the 3D velocity dispersion for isotopic turbulence, $v_{\textnormal{A,3D}}=B_{\textnormal{3D}}/\sqrt{\mu_0 \rho}$ is the 3D Alvénic velocity, $\mu_0$ is the permeability of vacuum. The estimated $\mathcal{M}_{\textnormal{A}}$ values are presented in \autoref{tab:b-field}. All the cores in Clump\,1 and Clump\,4 have shown $\mathcal{M}_{\textnormal{A}}\gtrsim1$, suggesting that turbulence is likely play an important role in these two clumps. Clump\,1-tail has a $\mathcal{M}_{\textnormal{A}}\sim0.7$, suggest that turbulence plays a less dominant role than magnetic field in this region.

\subsection{Virial Parameters}
To study the equilibrium states of the cores within each clump, we followed \cite{liujunhao2020} to calculate the virial parameter $\alpha_{\textnormal{k}+\textnormal{B}}$ by incorporating contributions from both the turbulent kinetic energy and the magnetic energy, which is given by
\begin{equation}
\alpha_{\textnormal{k}+\textnormal{B}}=\frac{M_{\textnormal{k}+\textnormal{B}}}{M},
\end{equation}
where $M_{\textnormal{k}+\textnormal{B}}$ is the total virial mass, including both the magnetic field virial mass and the kinetic mass, which can be estimated as 
\begin{equation}
M_{\textnormal{k}+\textnormal{B}}=\sqrt{M_{\textnormal{B}}^{2}+\left(\frac{M_{\textnormal{k}}}{2}\right)^2}+\frac{M_{\textnormal{k}}}{2}.
\end{equation}
The kinetic virial mass $M_{\textnormal{k}}$ and magnetic field virial mass $M_{\textnormal{B}}$ are given by
\begin{equation}
M_{\textnormal{k}}=\frac{3(5-2a)\sigma^{2}_{\textnormal{tot}}r}{(3-a)G},
\end{equation}
and
\begin{equation}
M_{\textnormal{B}}=\frac{\pi r^{2}B_{\textnormal{3D}}}{\sqrt{\frac{3(3-a)}{2(5-2a)}\mu_{0}\pi G}},
\end{equation}
respectively, where $r$ is the effective radius, $\sigma_{\textnormal{tot}}$ is the total gas velocity dispersion, and $a$ is the index of the density profile $\rho(r)\propto r^{-a}$. We adopted $a=1.6$, assuming a radial density profile \citep[e.g.,][]{pirogov2009} for all cores. We estimated the total gas velocity dispersion by using $\sigma_{\textnormal{tot}}=\sqrt{\sigma^{2}_{\textnormal{nt}}+\sigma^{2}_{\textnormal{th}}}$, in which $\sigma_{\textnormal{nt}}= \sqrt{\sigma^2_{\textnormal{obs}}-\frac{kT}{m_{\textnormal{obs}}} }$ is the non-thermal velocity dispersion, and $\sigma_{\textnormal{th}}= \sqrt{\frac{kT}{\mu_{\textnormal{p}} m_{\textnormal{H}}}}$ is the thermal velocity dispersion. Here $\sigma_{\textnormal{obs}}$ is the observed velocity dispersion, $m_{\textnormal{obs}}$ is the mass of the observed molecule (i.e., HN$^{13}$C, NH$_{3}$), $m_{\textnormal{H}}$ is the mass of the hydrogen atom, and $\mu_{\textnormal{p}}=2.33$ is the mean molecular weight per free particle \citep{kauffmann2008}.

The values of $\sigma_{\textnormal{tot}}$, $M_{\textnormal{k+B}}$, and $\alpha_{\textnormal{k+B}}$ are presented in \autoref{tab:b-field}.

\section{Discussion}

\subsection{Angular difference between multi-scale magnetic fields}\label{sec:multi-scale}
 
\begin{figure*}[ht]
\gridline{\hspace{-3.5\baselineskip}
          \fig{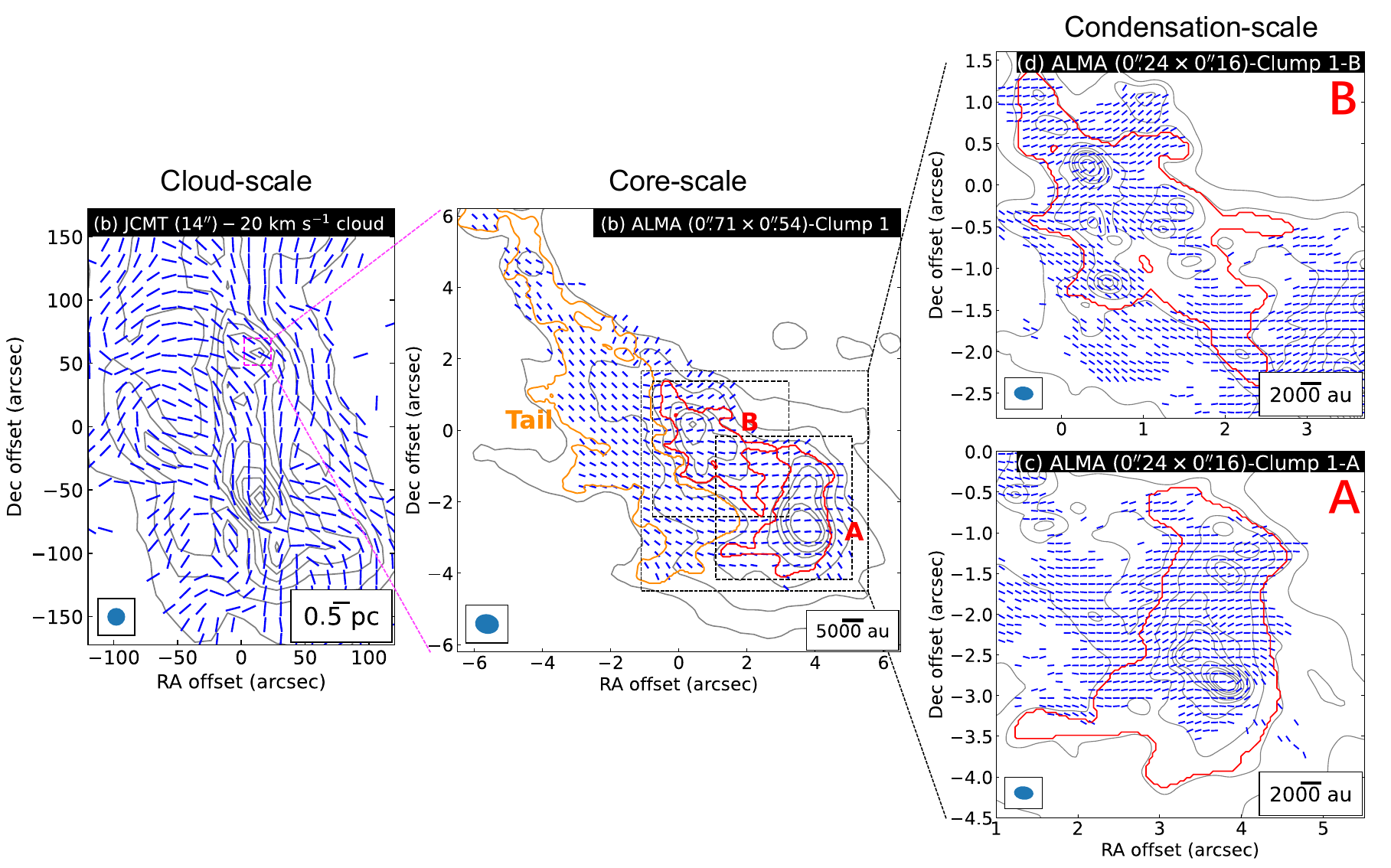}{1.06\textwidth}{}\hspace{-2.1\baselineskip}
          }
\vspace{-1\baselineskip}
\caption{The inferred magnetic field orientations at multi-scales for Clump\,1. The inferred magnetic field orientations (blue line segments) of Clump\,1 at a resolution of $\sim$14\arcsec revealed by JCMT data \citep{yang2025} in panel (a). The inferred magnetic field orientations (blue line segments) Clump\,1 revealed by the ALMA C-2 configuration in panels (b) with a resolution of $\sim$0\dotarcsec71$\times$0\dotarcsec54. The inferred magnetic field orientations (blue line segments) in Clump\,1 revealed by the ALMA C-5 configuration in panels (c) and (d) with a resolution of $\sim$0\dotarcsec24$\times$0\dotarcsec17. The synthesized beam is denoted by a filled blue ellipse at the bottom left corner in each panel. The grey contours shows the Stokes $I$ intensity in each panel. The contour levels are [5, 25, 45, 65, 85, 105, 125, 145, 165, 185] $\times$ $\sigma$ (1$\sigma$ = 27 mJy\,beam$^{-1}$) in panel (a), [5, 25, 50, 100, 150, 200, 300, 400] $\times$ $\sigma$ (1$\sigma$ = 0.22 mJy\,beam$^{-1}$) in panel (b), [25, 50, 100, 200, 300, 400, 600, 800, 1000, 1200] $\times$ $\sigma$ (1$\sigma$ = 32 $\mu$Jy\,beam$^{-1}$) in panels (c) and (d), respectively. The vectors in panels (b), (c), and (d) are selected using super-Nyquist sampling \cite[$\sim$3 pixel per beam;][]{hull2020} and set to have a uniform length.} 
\label{jcmt-alma-c1}         
\end{figure*}

\begin{figure*}[ht]
\gridline{\hspace{-4.5\baselineskip}
          \fig{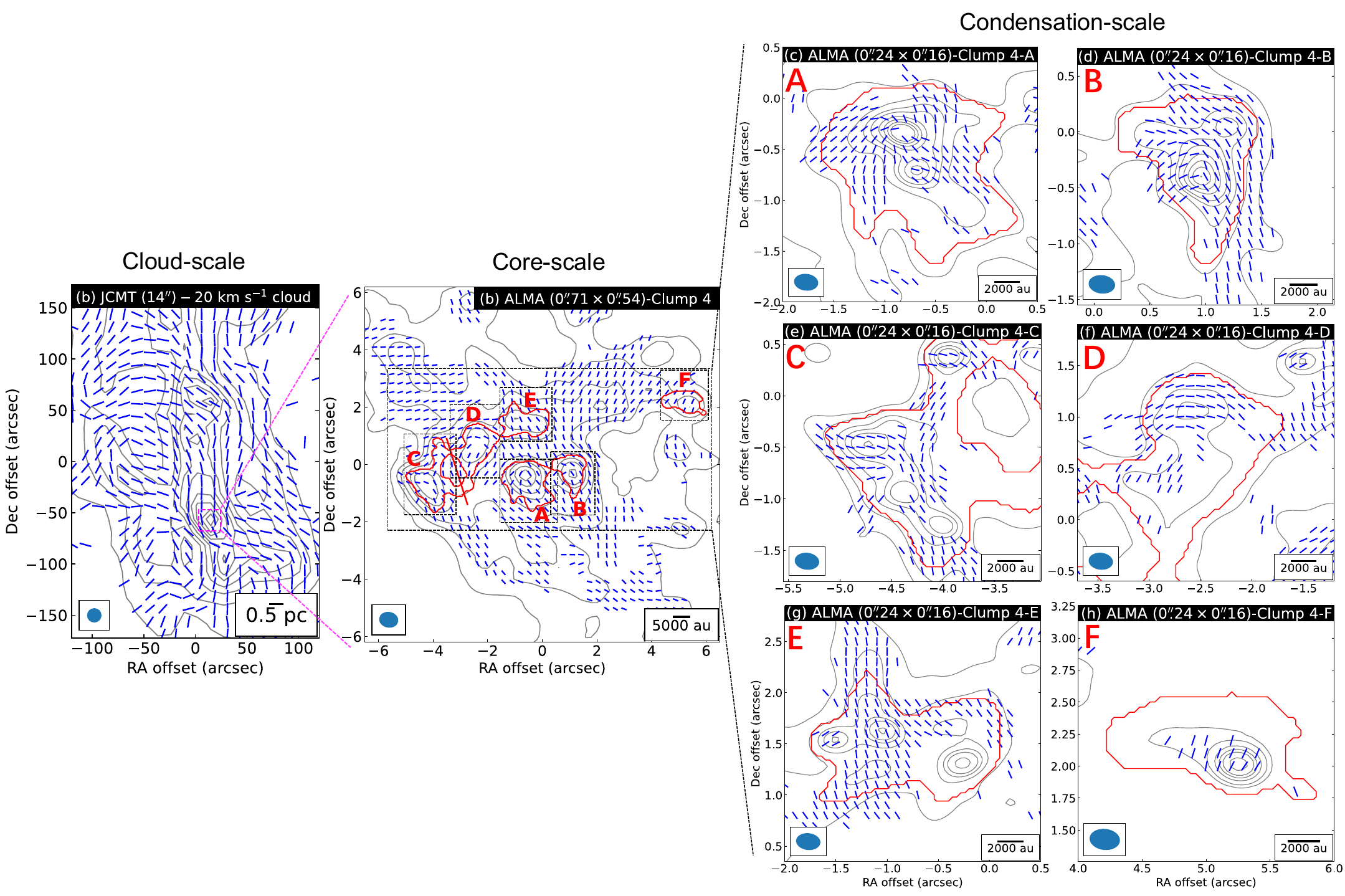}{1.05\textwidth}{}\hspace{-2.1\baselineskip}
          }
\vspace{-1\baselineskip}
\caption{The inferred magnetic field orientations at multi-scales for Clump\,4. The inferred magnetic field orientations (blue line segments) of Clump\,4 at a resolution of $\sim$14\arcsec revealed by JCMT data \citep{yang2025} in panel (a). The inferred magnetic field orientations (blue line segments) in Clump\,4 revealed by the ALMA C-2 configuration in panels (b) with a resolution of $\sim$0\dotarcsec71$\times$0\dotarcsec55. The inferred magnetic field orientations (blue line segments) in Clump\,4 revealed by the ALMA C-5 configuration in panels (c)$-$(h) with a resolution of $\sim$0\dotarcsec24$\times$0\dotarcsec16. The synthesized beam is denoted by a filled blue ellipse at the bottom left corner in each panel. The grey contours shows the Stokes $I$ intensity in each panel. The contour levels are [5, 25, 45, 65, 85, 105, 125, 145, 165, 185] $\times$ $\sigma$ (1$\sigma$ = 27 mJy\,beam$^{-1}$) in panel (a), [5, 25, 50, 100, 150, 200, 300, 400] $\times$ $\sigma$ (1$\sigma$ = 0.21 mJy\,beam$^{-1}$) in panel (b), [50, 100, 200, 300, 400, 600, 800, 1000, 1200] $\times$ $\sigma$ (1$\sigma$ = 22 $\mu$Jy\,beam$^{-1}$) in panels (c)$-$(h), respectively. The vectors in panels (b)$-$(h) are selected using super-Nyquist sampling \cite[$\sim$3 pixel per beam;][]{hull2020} and set to have a uniform length.}
\label{jcmt-alma-c4}         
\end{figure*}

\begin{figure*}[ht]
\gridline{\hspace{-1\baselineskip}
          \fig{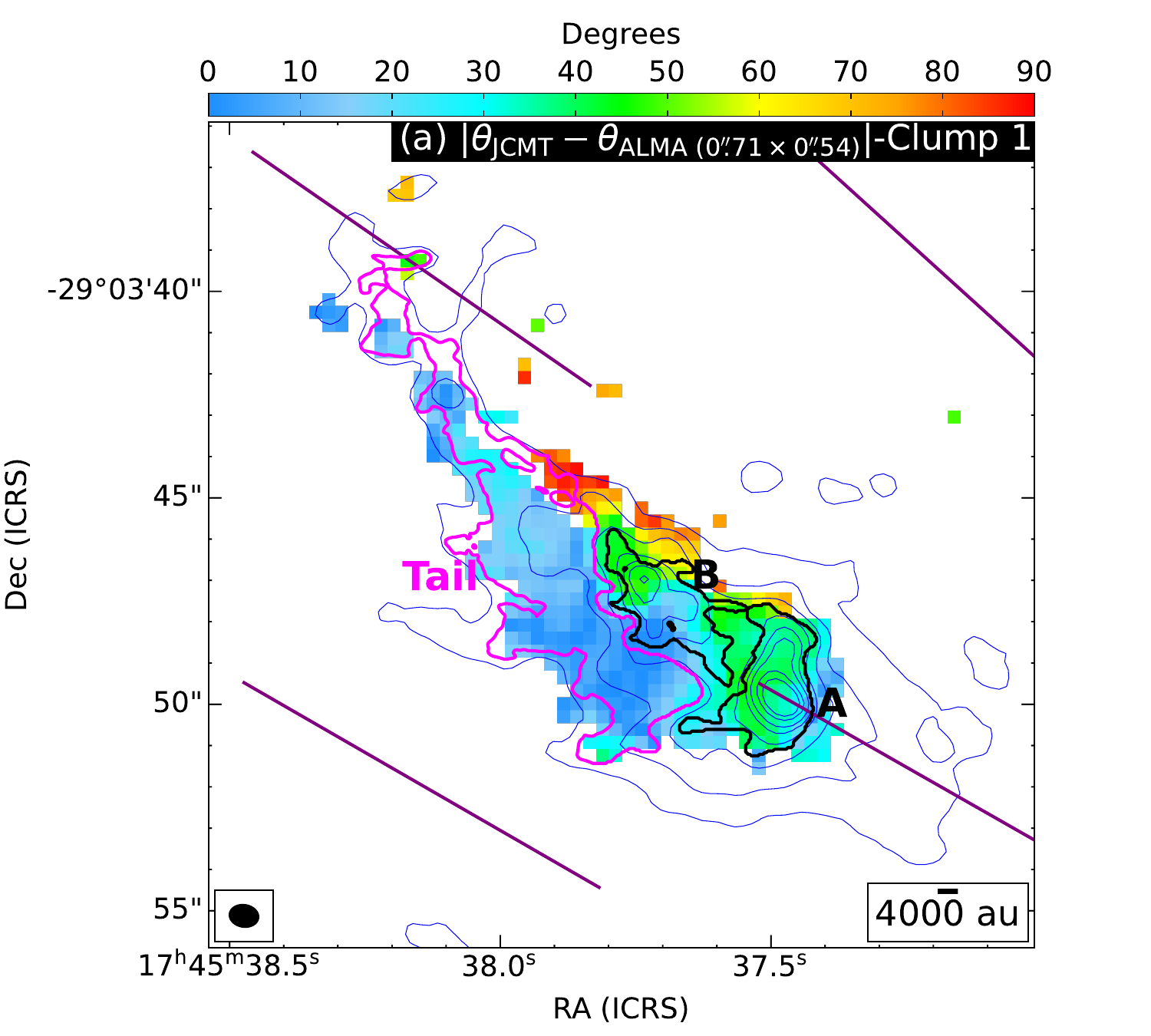}{0.5\textwidth}{}
          \hspace{-3\baselineskip}
          \fig{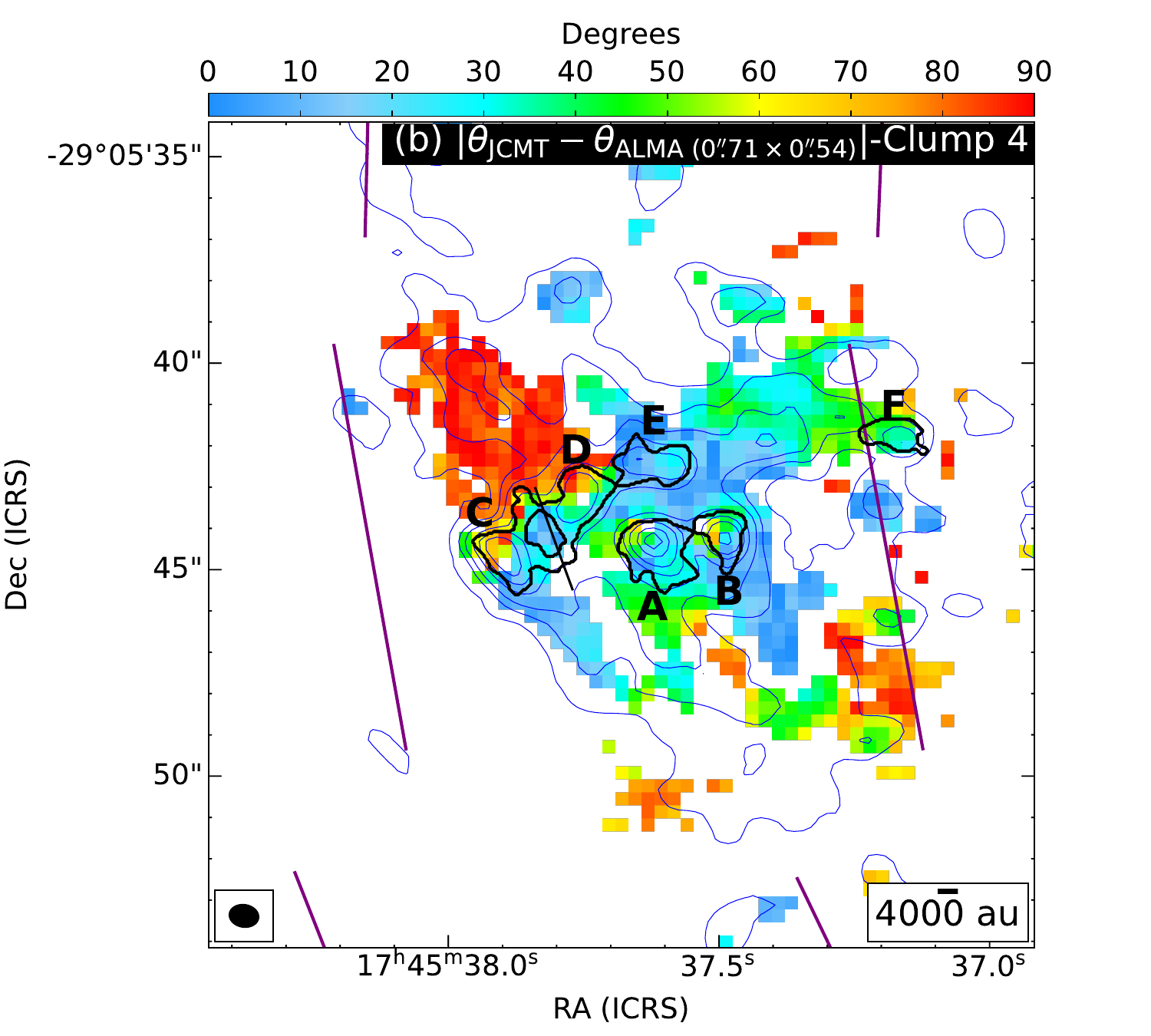}{0.5\textwidth}{}        \hspace{-2.1\baselineskip}
          }
\vspace{-1\baselineskip}
\caption{Absolute angular difference between the JCMT data and the ALMA C-2 configuration (resolution $\sim$0$\dotarcsec71\times0\dotarcsec54$) data shown in color for Clump\,1 and Clump\,4 in panels (a) and (b), respectively. The blue contours in each panel are the stokes $I$ intensity from the ALMA C-2 configuration data. The blue contours in panels (a) and (b) are identical to the black contours in \autoref{jcmt-alma-c1}\,(b) and \autoref{jcmt-alma-c4}\,(b), respectively. The cores enclosed by black solid contours in these two clumps are identified by \texttt{astrodendro}, and the magenta contour in panel (a) denotes Clump\,1-tail. The purple line segments in each panel show the inferred magnetic field orientation revealed by JCMT \citep{yang2025}.} 
\label{B_diff_angle}    
\end{figure*}

\begin{figure*}[ht]
\gridline{\hspace{-1\baselineskip}
          \fig{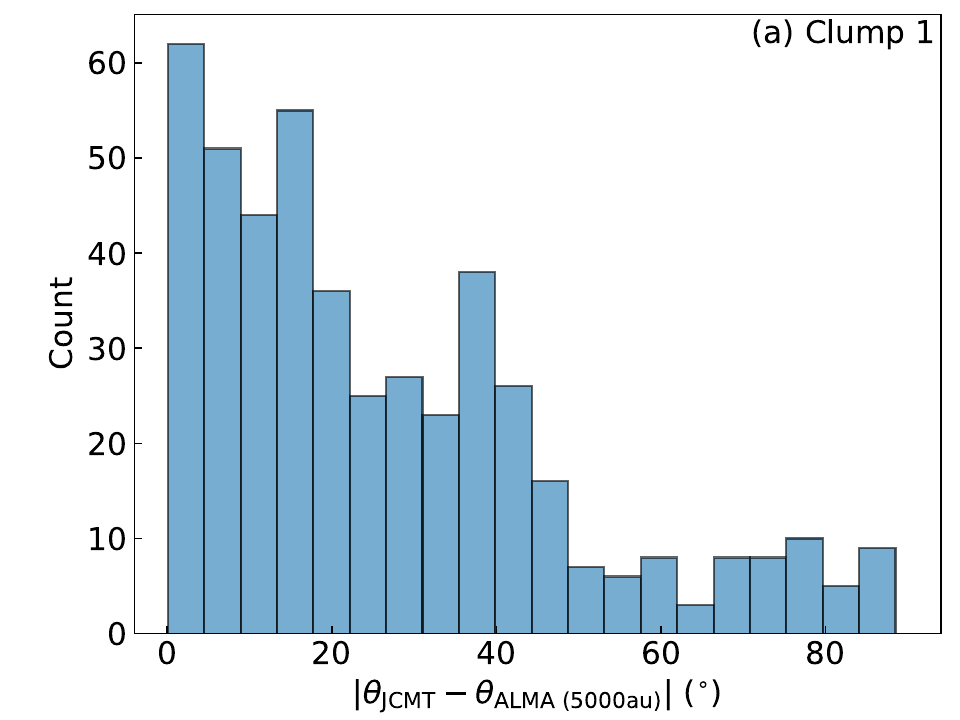}{0.45\textwidth}{}
          \hspace{-3\baselineskip}
          \fig{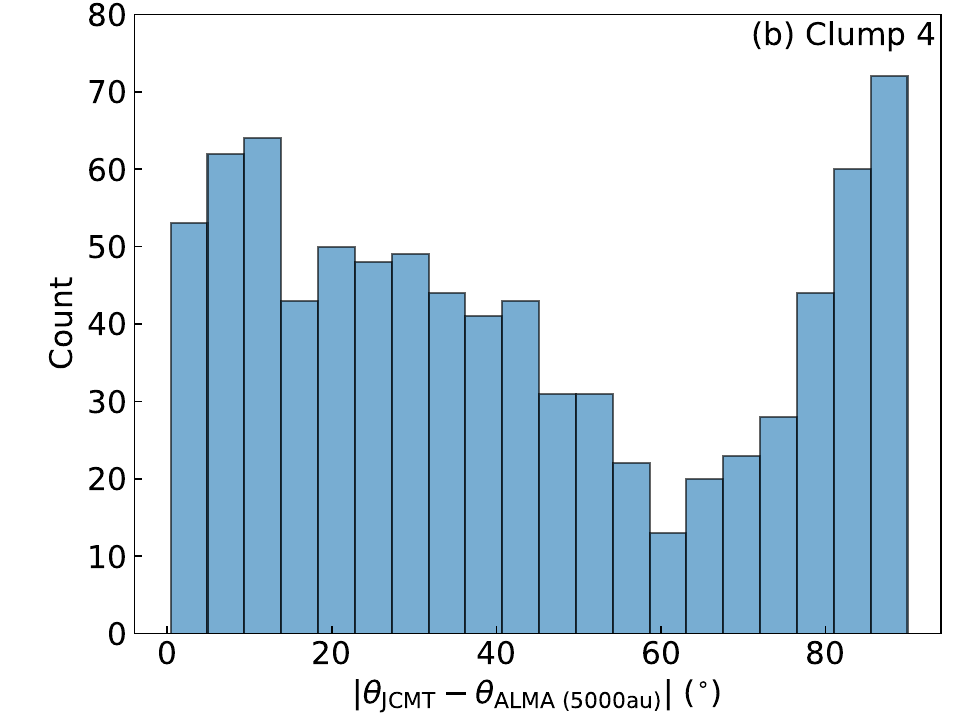}{0.45\textwidth}{}        \hspace{-2.1\baselineskip}
          }
\vspace{-1\baselineskip}
\gridline{\hspace{-1\baselineskip}
          \fig{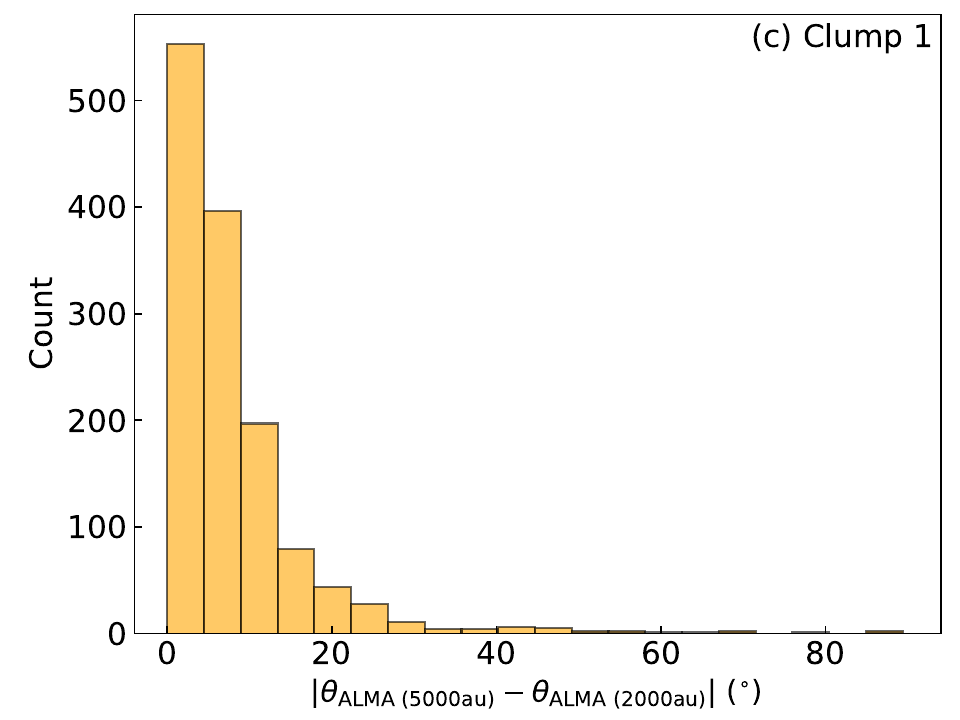}{0.45\textwidth}{}
          \hspace{-3\baselineskip}
          \fig{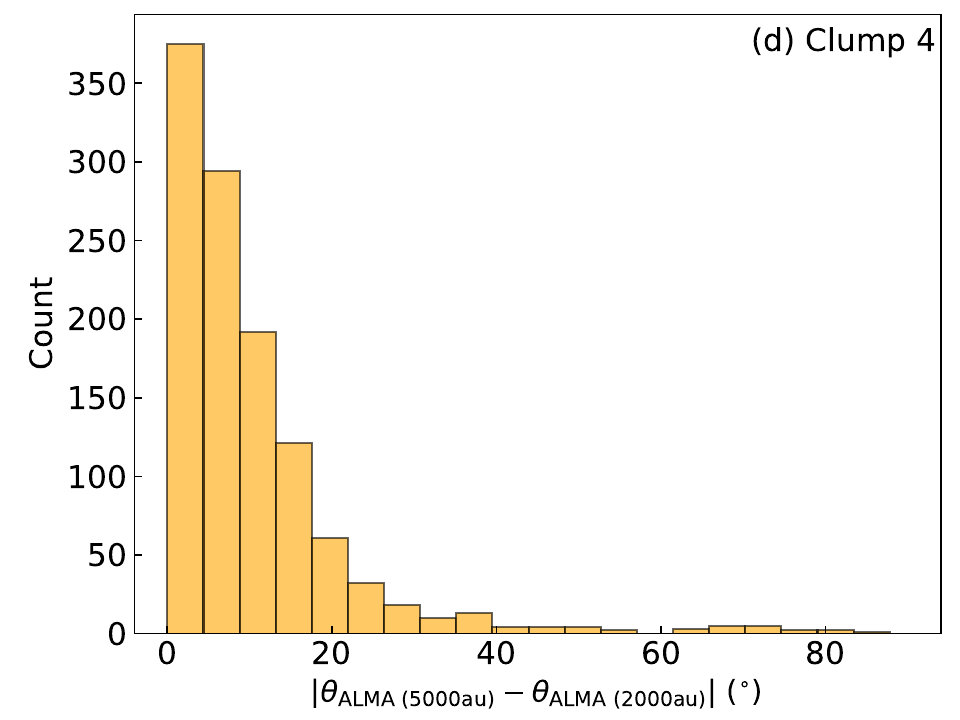}{0.45\textwidth}{}        \hspace{-2.1\baselineskip}
          }
\vspace{-1\baselineskip}
\caption{Top panels: Histogram of the absolute angular difference between the JCMT data and the ALMA C-2 configuration (resolution $\sim$5000 au) for Clump\,1 and Clump\,4 in panels (a) and (b), respectively. Bottom panels: Histogram of the absolute angular difference between the ALMA C-2 (resolution $\sim$5000 au) and C-5 configurations (resolution $\sim$2000 au) for Clump\,1 and Clump\,4 in panels (c) and (d), respectively.} 
\label{ro-jcmt-alma}    
\end{figure*}

In this section, we use the JCMT data \citep{yang2025}, ALMA C-2 configuration data, and ALMA C-5 configuration data to compare the magnetic field structures in the Clump\,1 and Clump\,4 at three different resolutions: $\theta\sim14\arcsec$ \citep[JCMT;][]{yang2025}, $\theta\sim0\dotarcsec71\times0\dotarcsec54$ (ALMA C-2), $\theta\sim0\dotarcsec24\times0\dotarcsec16$ (ALMA C-5). Those resolutions correspond to 0.6\,pc, 5000\,au, and 2000\,au in linear scales, respectively. These scales in this work are referred to as the cloud-scale, the core-scale, and the condensation-scale, respectively. We iterate over each pixel to calculate the distance between all pairs of vectors from these three data sets. We then compute the absolute difference between the angles from one data set with the angles from the nearest pixels in the other data set. The inferred magnetic field orientations at multi-scales for Clump\,1 and Clump\,4 are shown in the \autoref{jcmt-alma-c1} and \autoref{jcmt-alma-c4}, respectively.

\autoref{jcmt-alma-c1}\,(a) shows the magnetic field orientations at cloud-scale in Clump\,1 aligned orderly along a northeast-southwest direction. \autoref{jcmt-alma-c1}\,(b) shows the magnetic field at core-scale. The magnetic field orientations of Clump\,1-tail are consistent with those at cloud-scale, which align parallel with its major axis with a mean difference of $\sim$15$^{\circ}$ (see \autoref{B_diff_angle}\,(a)). The mean difference for all cores and Clump 1-tail between these two scales is calculated as the average value within the core masks (black contours in \autoref{B_diff_angle}) and the Clump 1-tail mask (magenta contour \autoref{B_diff_angle}). The magnetic field orientations of Clump\,1-A are nearly parallel to its minor axis, which align along the east-west direction. Clump\,1-B shows a similar magnetic field orientation to Clump\,1-A. The magnetic field orientations of Clump\,1-A and Clump\,1-B at core-scale show the deviations of $\sim$35$^{\circ}$ and $\sim$29$^{\circ}$ from those at cloud-scale, respectively (see \autoref{B_diff_angle}\,(a)). At core-scale, the magnetic field orientations transit from being perpendicular to the continuum contours in the two cores to becoming parallel to the contours in Clump\,1-tail. \Cref{jcmt-alma-c1}\,(c) and (d) show that the two cores detected at core-scale are resolved into multiple compact condensations. The vectors are mostly detected from the dense regions at condensation-scale. Since the diffuse Clump\,1-tail is only detected with a limited number of vectors, we only present the zoomed-in images of Clump\,1-A and Clump\,1-B. \autoref{ro-jcmt-alma}\,(c) shows the histogram distributes in a right-skewed triangle distribution, which means that the orientations between these two scales are generally consistent. There are only 16\% of the vectors showing an angular difference of $\ge$45$^{\circ}$. 

At the cloud-scale, the magnetic field plays a significant role in providing support against gravitational collapse \citep{yang2025}. At the core-scale, the orientation in the diffuse Clump\,1-tail shows consistence with the cloud-scale magnetic field orientation, as the magnetic field could effectively resist the gravitational collapse. On the other hand, the dense Clump\,1-A and Clump\,1-B show different orientations compared to those of the cloud-scale, as the density structures are mainly shaped by gravity in the high density regions \citep{koch2018}. At the condensation-scale, the magnetic field orientations are predominately consistent with those at the core-scale, which suggests that gravity plays the important role in shaping the density structure in the dense regions at both scales. In addition, the sub-virial state also suggested that gravitational forces likely dominate gas dynamics at this scale. However, at condensation-scale, star formation activities \citep{liujunhao2023} could also be a dominant force as explained in \autoref{sec:B-LG}.

\autoref{jcmt-alma-c4}\,(a) shows that the magnetic field orientations at cloud-scale in Clump\,4 primarily distribute along a north-south direction. \autoref{jcmt-alma-c4}\,(b) shows the magnetic field orientations at core-scale. Even in diffuse regions, the magnetic field is oriented in different directions. Six cores were detected, and all of them show deviations in magnetic field orientations from that at the cloud-scale, with Clump\,4-A, Clump\,4-B, Clump\,4-C, Clump\,4-D, Clump\,4-E and Clump\,4-F show the deviations of $\sim$31$^{\circ}$, $\sim$30$^{\circ}$, $\sim$42$^{\circ}$, $\sim$48$^{\circ}$, $\sim$16$^{\circ}$, and $\sim$41$^{\circ}$, respectively (see \autoref{B_diff_angle}\,(b)). The discrepancies between these two scales are prominent in Clump\,4 as evidenced by the binominal distribution in \autoref{ro-jcmt-alma}\,(b). \autoref{jcmt-alma-c4}\,(c)$-$(h) show the magnetic field at condensation-scale, which reveal that the magnetic field orientations are generally consistent between the core- and condensation-scales. The complex magnetic field orientation at the condensation-scale may be affected by the H\textsc{ii} region nearby as shown in \autoref{cband}\,(b). However, we will not delve into this topic further in this work.

The magnetic orientation changes in Clump\,4 at different scales generally follows those observed in Clump\,1. Discrepancies in the magnetic field orientation occur between the cloud- and core-scales for both Clump\,1 and Clump\,4. This could happen when the magnetic field dominates at the cloud-scale, whereas the gravity becomes the predominant force at the core-scale. Such discrepancies also suggest a transition of the magnetic critical state between these two scales \citep{liujh2024}. In contrast, the magnetic field orientations are mostly consistent between the core- and condensation-scales for both Clump\,1 and Clump\,4, which could be attributed to the dominance of gravitational forces at scales smaller than 5000\,au. 

\subsection{Relative Orientation between Magnetic Field versus Local Gravity}\label{sec:B-LG}
In this section, we follow the method introduced by \cite{liujunhao2023} to study the magnetic field and density structures in Clump\,1 and Clump\,4. The column density gradient and velocity gradient were derived using a 3$\times$3 Sobel kernel \citep{soler2013,liujunhao2023}. The local gravity gradient was derived from the column density map using the standard formula of gravity \citep[][]{koch2012a,liujunhao2020,he2023}. The details of calculation and uncertainty estimations can be found in \cite{liujunhao2023}. The angle difference between these orientations can be characterized by the alignment measure \citep[AM;][]{casanova2017,lazarian2018,liujunhao2023}, which is given by
\begin{equation}
\textnormal{AM}=\langle\cos(2\phi)\rangle,
\end{equation}
where $\phi= \left| \theta_{\textnormal{1}}-\theta_{\textnormal{2}}\right| $ represents the angle difference in the range of 0$^{\circ}$$-$90$^{\circ}$ between the orientations of any two of the magnetic field ($\theta_{\textnormal{B}}$), gas column density gradient ($\theta_{\textnormal{NG}}$), local gravity gradient ($\theta_{\textnormal{LG}}$), and velocity gradient ($\theta_{\textnormal{VG}}$). The AM is ranged from $-$1 to 1, with $-$1 indicating perpendicular alignment and 1 indicating parallel alignment. The AM is calculated in different $N_{\textnormal{H}_2}$ bins and each bin contains equal number of pixels. Each $N_{\textnormal{H}_2}$ bin in both Clump\,1 and Clump\,4 contains at least 30 independent measurements. The uncertainty of AM is estimated as $\delta\textnormal{AM}=\sqrt{(\langle(\cos(2\phi))^{2}\rangle-\textnormal{AM}^{2}+\Sigma^{n^{\prime}}_{i}(2\sin(2\phi_{i})\delta\phi_{i})^{2})/n^{\prime}}$, where $n^{\prime}$ is the number of data points, $\phi_{i}$ and $\delta\phi_{i}$ are the angle difference and its uncertainty, respectively, associated with the $i$-th data point \citep{liujunhao2023}. Among the relative orientations ($\phi_\textnormal{B}^\textnormal{LG}$, $\phi_\textnormal{B}^\textnormal{NG}$,$\phi_\textnormal{NG}^\textnormal{LG}$, $\phi_\textnormal{VG}^\textnormal{B}$, $\phi_\textnormal{VG}^\textnormal{LG}$, and $\phi_\textnormal{VG}^\textnormal{NG}$), except for $\phi_\textnormal{B}^\textnormal{LG}$, all the other orientations do not show a prominent trend, and we refrain from further discussing them.

\begin{figure*}[ht]
\gridline{\hspace{-5\baselineskip}
          \fig{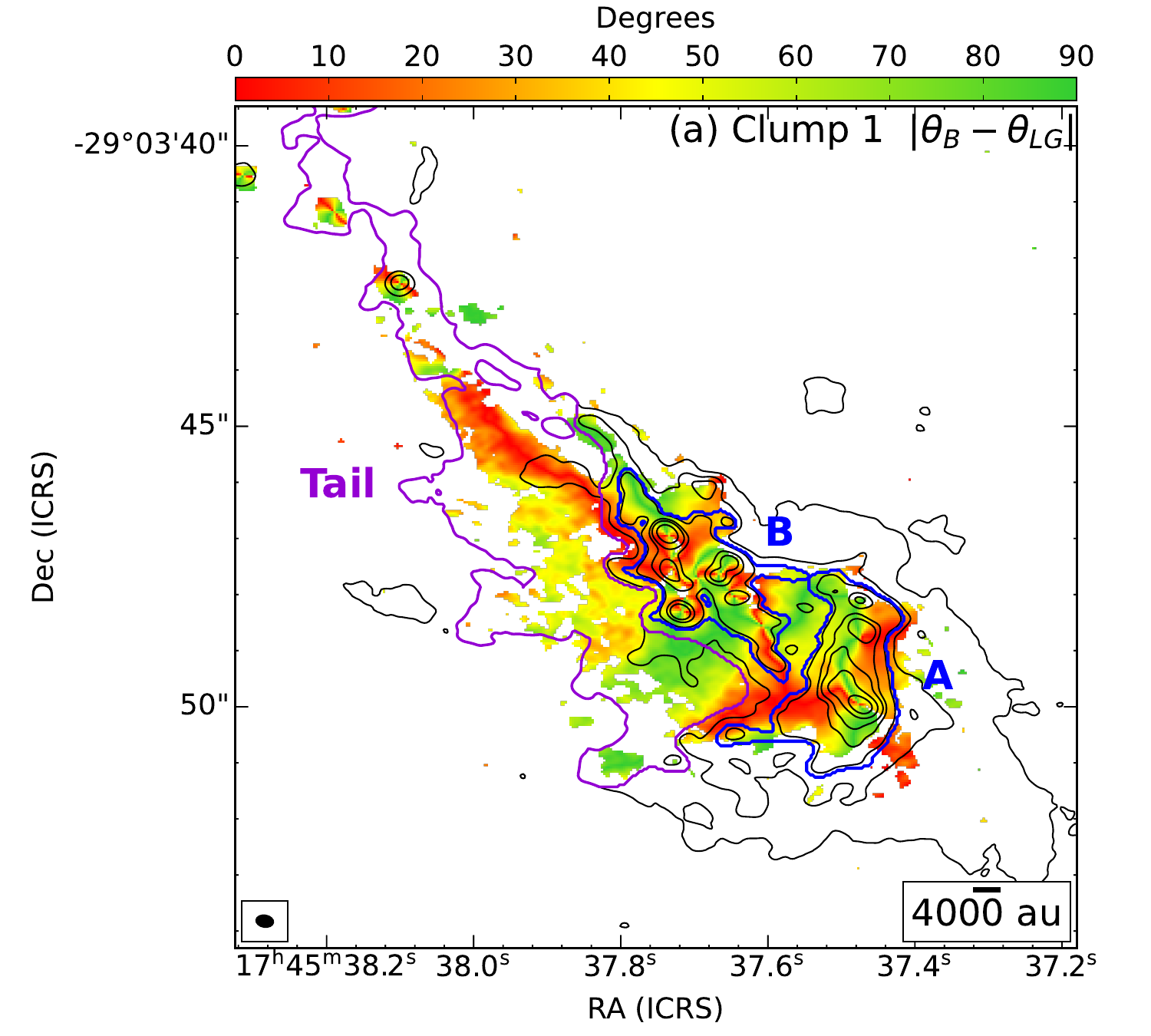}{0.55\textwidth}{}
          {}\hspace{-4\baselineskip}
          \fig{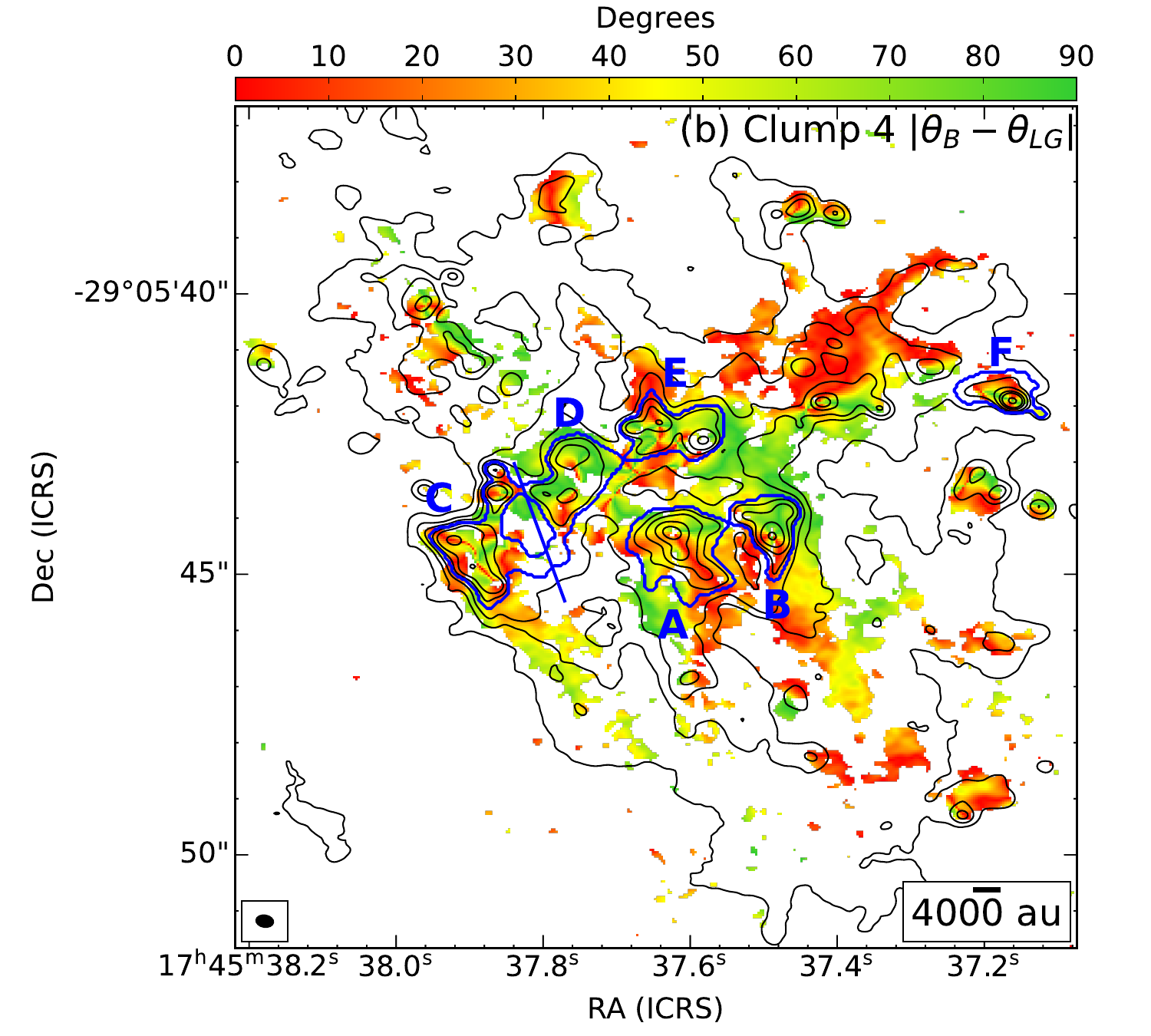}{0.55\textwidth}{}
          {}\hspace{-2.1\baselineskip}
          }
\vspace{-1\baselineskip}
\caption{Relative orientation between the inferred magnetic field ($\theta_{\textnormal{B}}$) and local gravity ($\theta_{\textnormal{LG}}$) in color overlaid with the Stokes $I$ emission in black contours. The black contours are [5, 25, 50, 100, 175, 350, 700] $\times$ $\sigma$ (1$\sigma$ = 65 $\mu$Jy\,beam$^{-1}$) in panel (a) and [5, 25, 50, 100, 150, 300, 600] $\times$ $\sigma$ (1$\sigma$ = 48 $\mu$Jy\,beam$^{-1}$) in panel (b). The cores enclosed by blue solid contours in these two clumps are identified by \texttt{astrodendro}, and purple contour in panel (a) denotes Clump\,1-tail. The synthesized beam size is denoted by the filled black ellipse in the bottom left corner.} 
\label{delta_B-LG}          
\end{figure*}

\begin{figure*}[ht]
\gridline{\hspace{-5\baselineskip}
          \fig{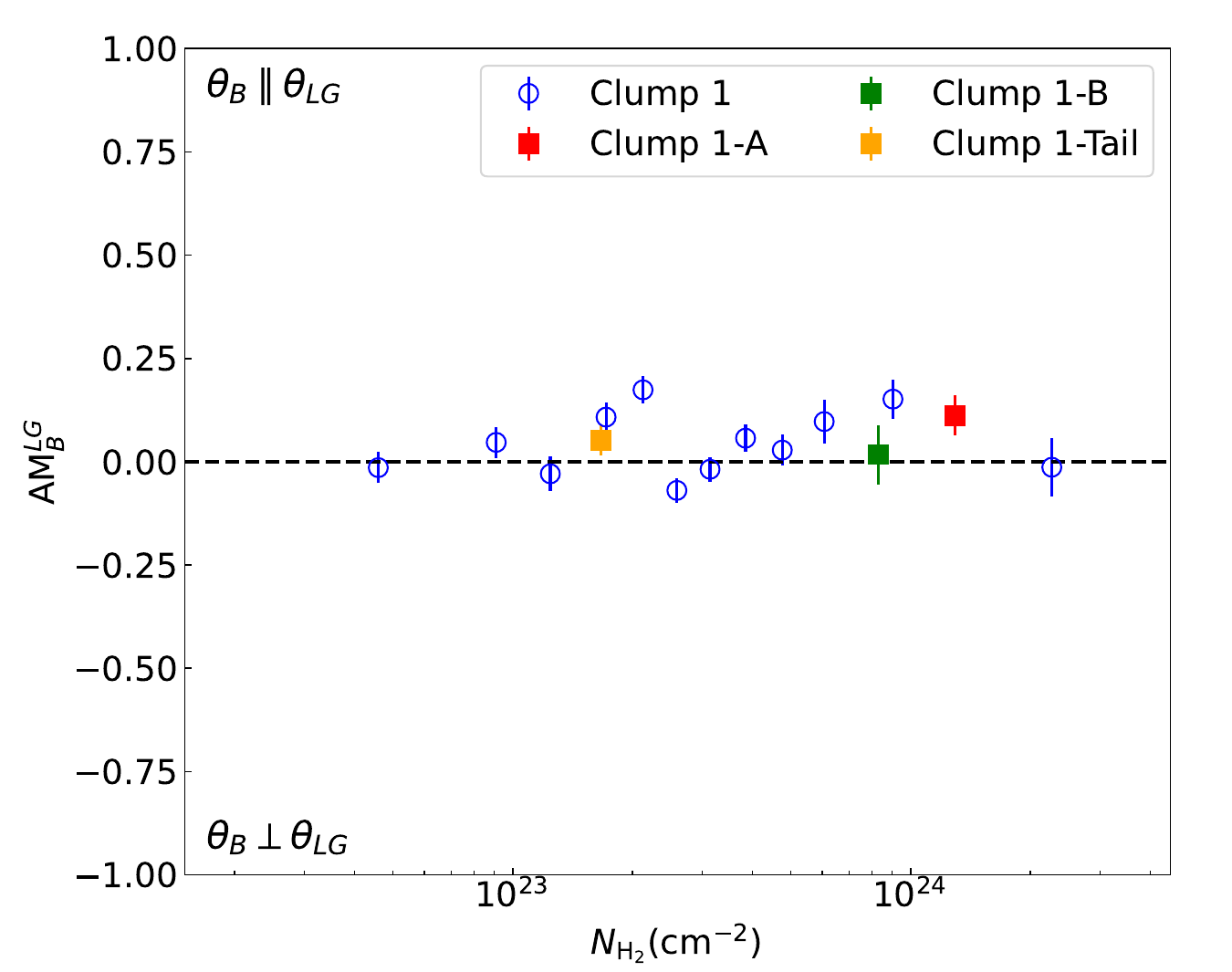}{0.5\textwidth}{(a)}{}\hspace{-2\baselineskip}
          \fig{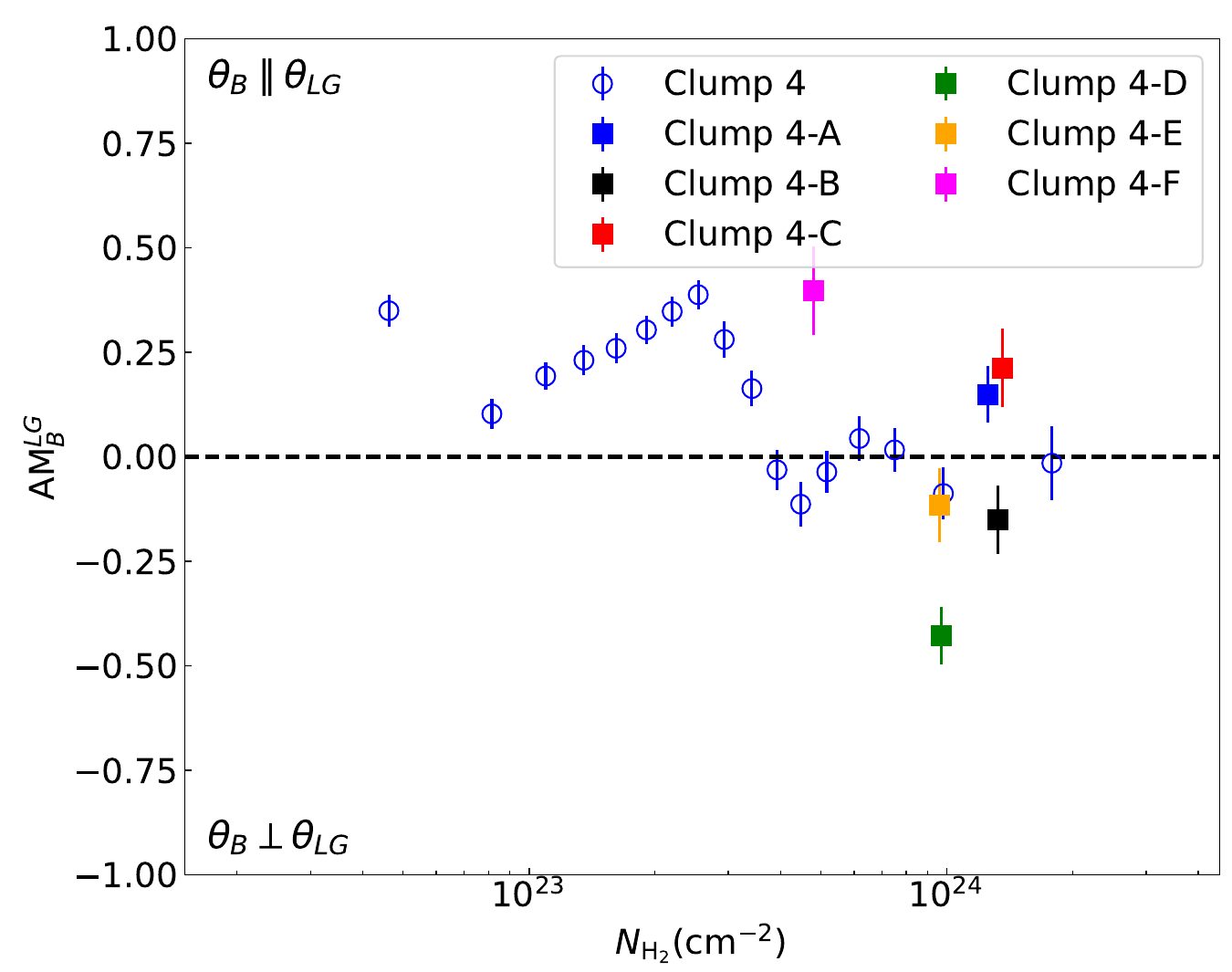}{0.5\textwidth}{(b)}      {}\hspace{-2.1\baselineskip}}
\vspace{-1\baselineskip}
\caption{Relative orientations between $\theta_{\textnormal{B}}$ and $\theta_{\textnormal{LG}}$ for (a) Clump\,1 and (b) Clump\,4. The empty blue circles denote $N_{\textnormal{H}_{2}}$ bins obtain for the whole clump and they are divided to 12 bins for (a) Clump\,1 and 18 bins (b) Clump\,4 ($>30$ independent measurements in each bin) in each plot. Different cores in each plot are color-coded with squares. Each core and tail in Clump\,1 is divided into 1 $N_{\textnormal{H}_{2}}$ bins ($>45$ independent measurements in each bin). Each core in Clump\,4 is divided into 1 $N_{\textnormal{H}_{2}}$ bins ($>6$ independent measurements in each bin). The black dashed line at $\textnormal{AM}=0$ denotes the random alignment. $\textnormal{AM}>0$ is defined as the parallel alignment, and $\textnormal{AM}<0$ is defined as the perpendicular alignment.} 
\label{am_B-LG}          
\end{figure*}

The relative orientation between the magnetic field and local gravity ($\phi_\textnormal{B}^\textnormal{LG}$) quantifies the efficiency of gravity in structuring magnetic field lines and aligning them with gravitational pull \citep{koch2012a,liujunhao2023}. The low values of $\phi_\textnormal{B}^\textnormal{LG}$ indicate that the magnetic field and gravity are mostly aligned with a small angular difference in orientation. In this case, the magnetic field may be ineffective at providing resistance against gravity. The high values of $\phi_\textnormal{B}^\textnormal{LG}$ indicate that the magnetic field and gravity are misaligned with a large angular difference in orientation, which could allow the magnetic field to provide resistance against gravity \citep{koch2018}. \Cref{delta_B-LG}\,(a) and (b) show the spatial distribution of $\phi_\textnormal{B}^\textnormal{LG}$ for Clump\,1 and Clump\,4, respectively. Clump\,1-tail shows low $\phi_\textnormal{B}^\textnormal{LG}$ values, while Clump\,1-A and Clump\,1-B show both low and high $\phi_\textnormal{B}^\textnormal{LG}$ values. All cores in Clump\,4 show both low and high $\phi_\textnormal{B}^\textnormal{LG}$ values. The tangential fan-like structure observed near the peak of the continuum emission reported by \cite{koch2018} and \cite{liujunhao2023} has also been observed toward the peaks of the condensations of Clump\,1 and Clump\,4. The areas of the fan-like structures showing low $\phi_\textnormal{B}^\textnormal{LG}$ values suggest that gravitational collapse can occur without much of magnetic resistance \citep{koch2018,liujunhao2023}, where the magnetic field acts as a channel to transfer materials \citep{koch2018}. The Clump\,1-tail shows mostly the low $\phi_\textnormal{B}^\textnormal{LG}$ values close to 0$^{\circ}$, which indicates a predominantly parallel alignment. The parallel alignment observed in Clump\,1-tail suggests gravity becomes increasingly dominant in shaping the magnetic field structure in this diffuse region.

\autoref{am_B-LG}\,(a) shows that the AM$_\textnormal{B}^\textnormal{LG}$ values in both the entire clump and the cores of Clump\,1 suggest a random alignment. The random alignments suggest a trans-critical state or the distortions of magnetic field may arise from sources other than gravity. This is because gravity-induced distortions typically show as parallel alignments instead of the random alignments \citep[e.g.,][]{koch2018,zhang2025}. The potential sources contributing to the random alignments may include star formation activities such as outflows and converging flows \citep{liujunhao2023}.

\autoref{am_B-LG}\,(b) shows that the AM$_\textnormal{B}^\textnormal{LG}$ values for Clump\,4 generally increase with increasing $N_{\textnormal{H}_{2}}$, which suggests that a preferentially more parallel alignment transits to a random alignment. The alignments suggest that magnetic field could work effectively in Clump\,4-D, whereas Clump\,4-F is overwhelmed by the gravity without much resistance from magnetic field \citep{koch2018}. The gravity likely dominates the Clump\,4-A and Clump\,4-C with small contribution from magnetic field and star formation activities. The alignments in Clump\,4-B and Clump\,4-E suggest that magnetic field could work effectively with some impact from gravity and the star formation activities \citep{liujunhao2023}. Overall, in both clumps, the alignment we observed is inconsistent with the preferentially parallel configuration observed across the CMZ \citep{pan2025}. It is also inconsistent with the alignment observed toward the 17 massive protostellar cluster-forming clumps in the Galactic disk, where the magnetic field and column density show parallel alignment at higher column densities \citep{zhang2025}. The former suggests that most dense gas structures in the CMZ are stable against gravitational collapse, supported by strong turbulence and magnetic fields \citep{pan2025}. The latter indicates that gravitational collapse occurs on envelope scales as the magnetic field is dragged inward by infalling motions in magnetically supercritical gas \citep{zhang2025}. In contrast, our results showing a random alignment suggest that the these two clumps are predominantly trans-critical, or that the magnetic field is significantly influenced by star formation feedback \citep{liujunhao2023}.

\subsection{Magnetic tension force ($F_\textnormal{B}$) versus the Gravitational force ($F_\textnormal{G}$) between the cores}

\begin{figure*}[ht]
\gridline{\hspace{-1\baselineskip}
          \fig{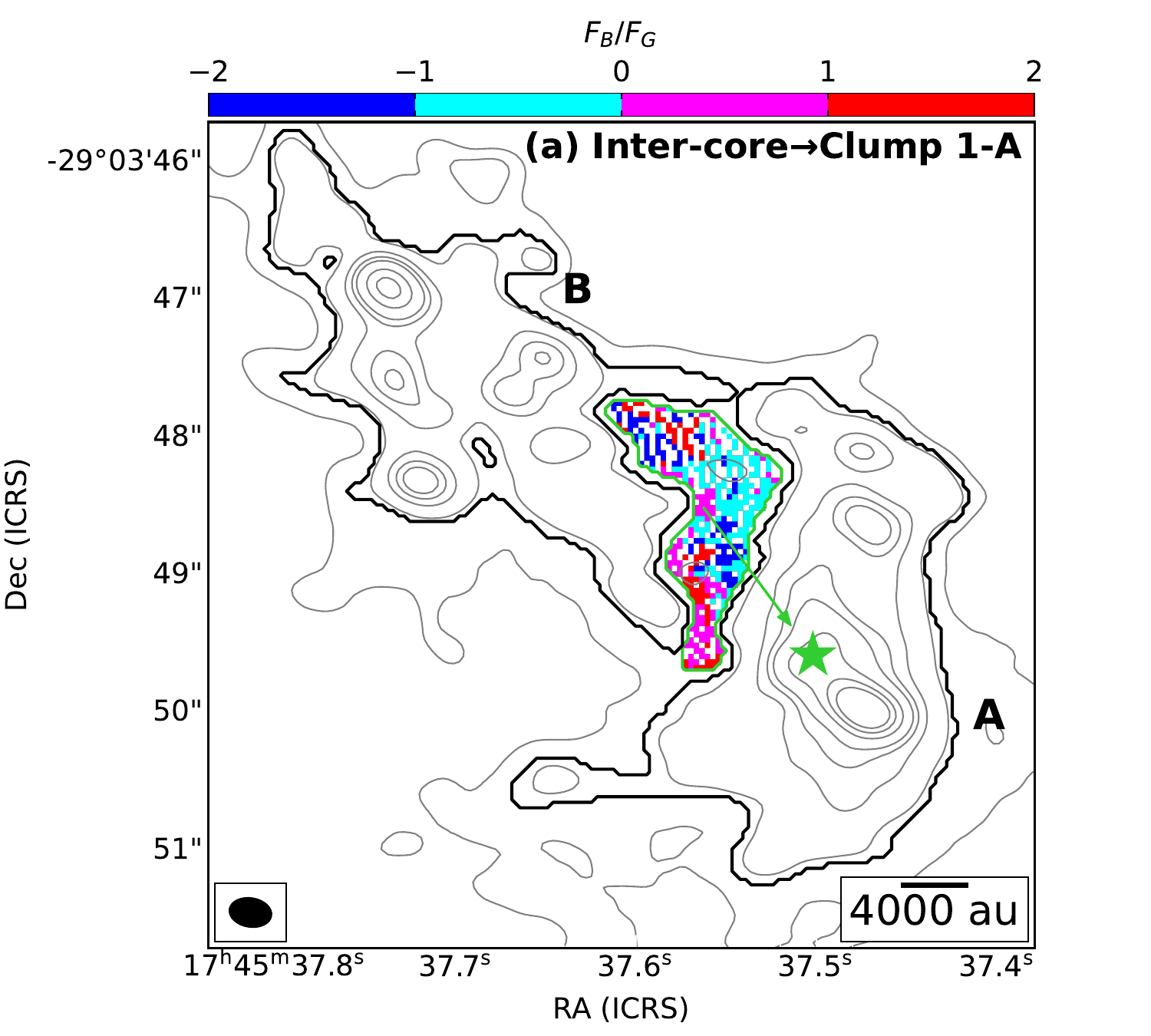}{0.5\textwidth}{}
          \hspace{-3\baselineskip}
          \fig{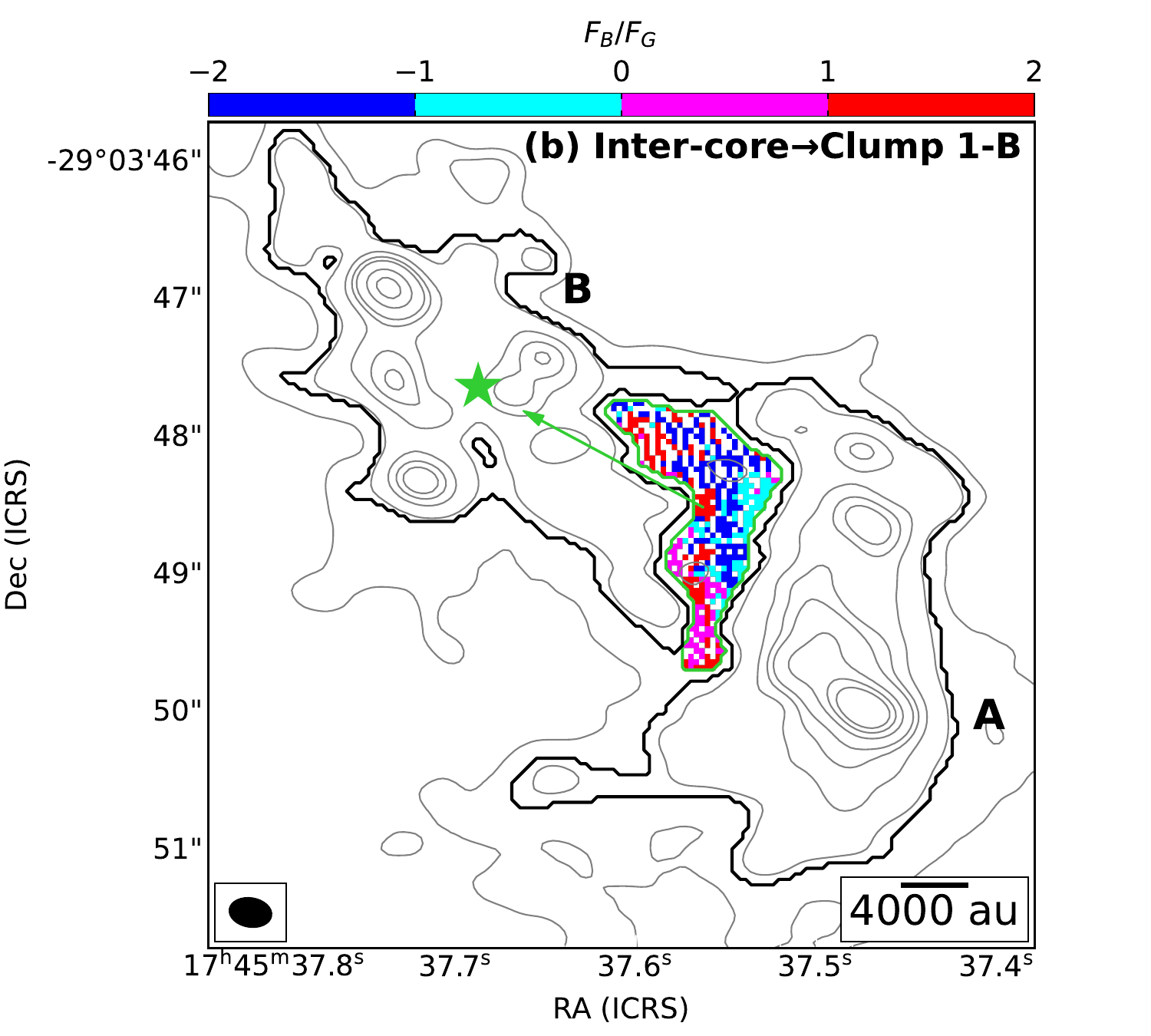}{0.5\textwidth}{}        \hspace{-2.1\baselineskip}
          }
\vspace{-2\baselineskip}
\gridline{\hspace{-1\baselineskip}
          \fig{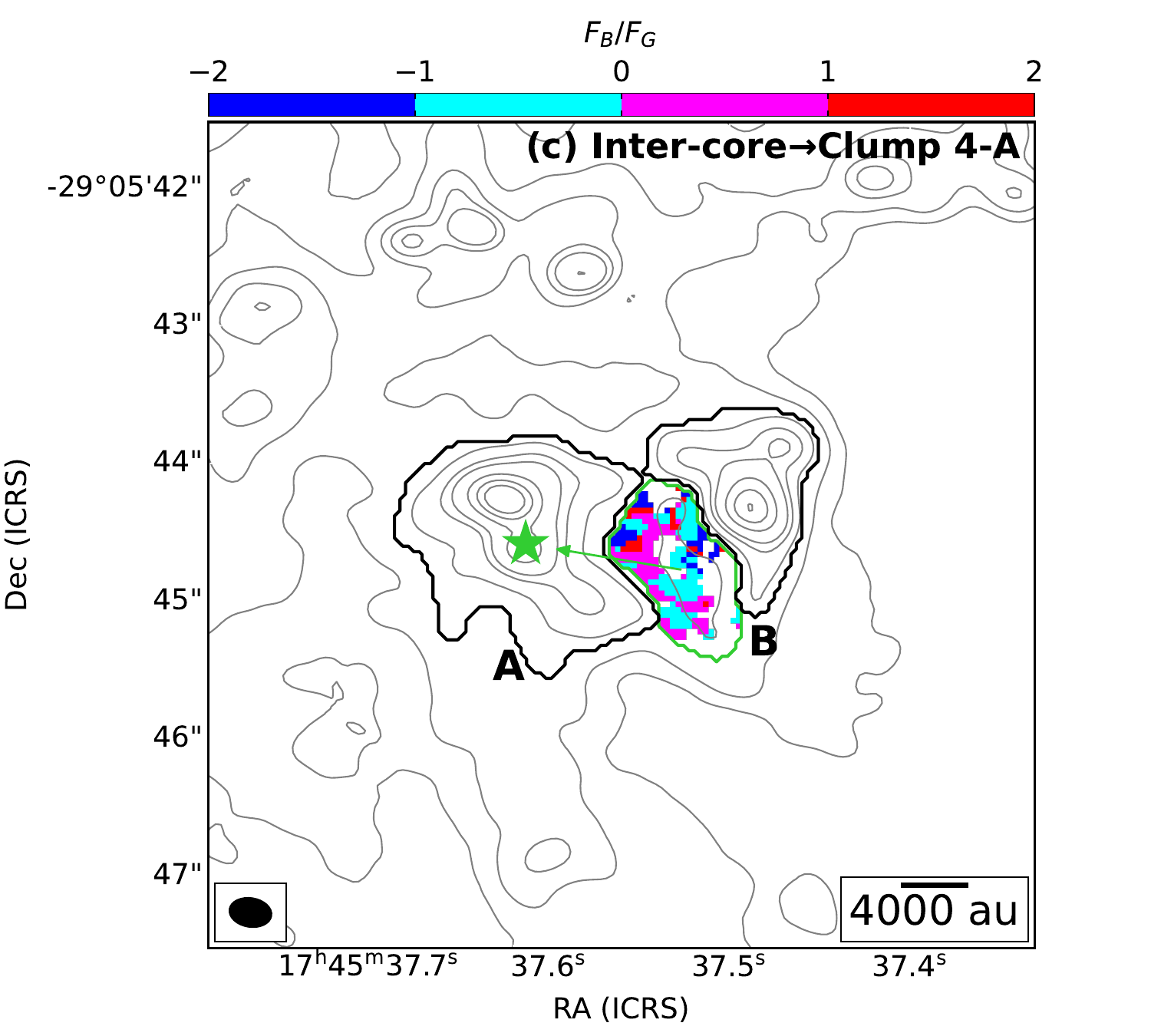}{0.5\textwidth}{}
          \hspace{-3\baselineskip}
          \fig{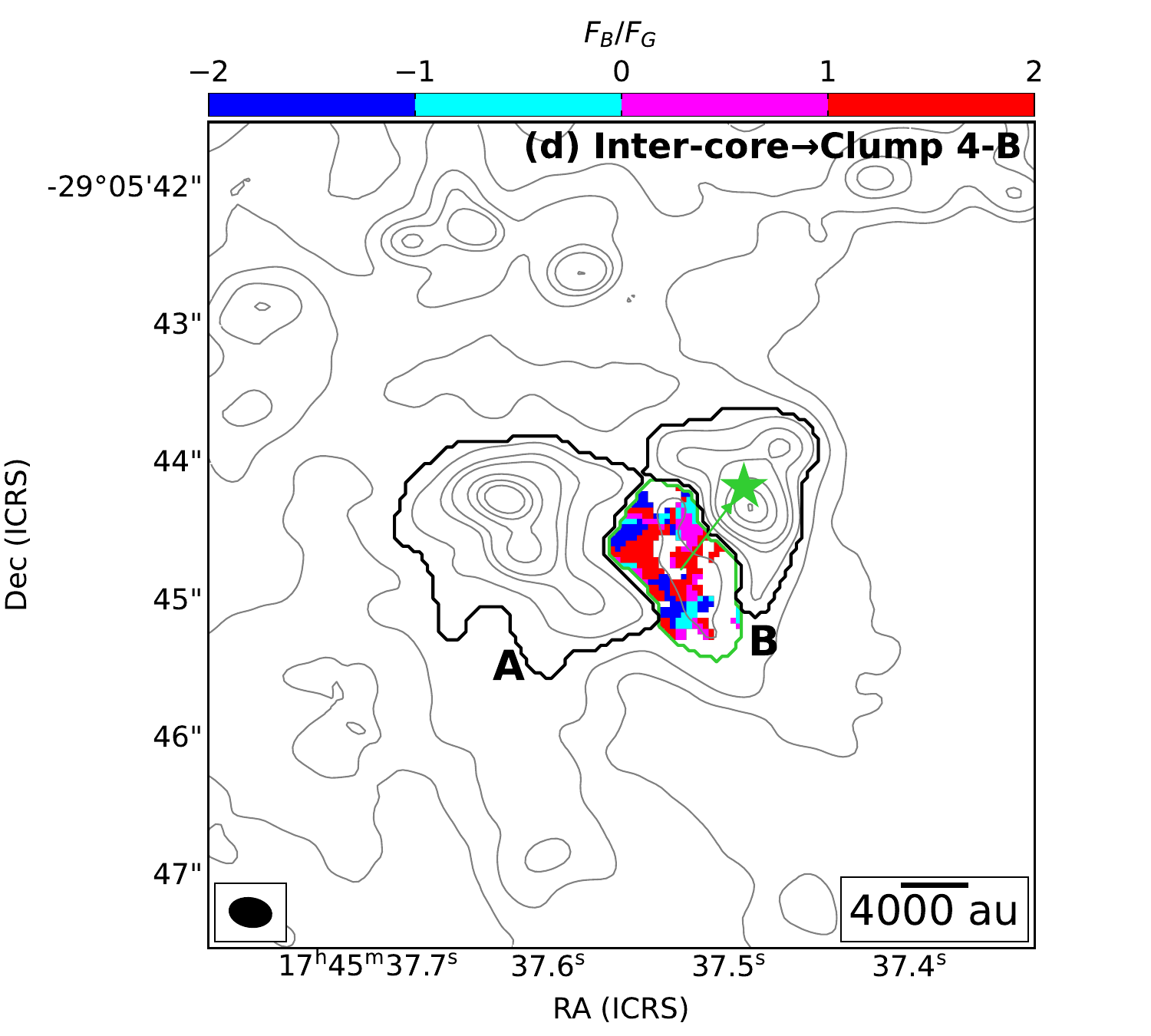}{0.5\textwidth}{}        \hspace{-2.1\baselineskip}
          }
\vspace{-1\baselineskip}
\caption{Top: $F_\textnormal{B}/F_\textnormal{G}$ ratios for the inter-core in Clump\,1 (color). The gray contours are the Stokes $I$ emission, and the contour levels are [25, 50, 100, 150, 200, 350, 500, 650] $\times$ $\sigma$ (1$\sigma$ = 65 $\mu$Jy\,beam$^{-1}$). Bottom: The $F_\textnormal{B}/F_\textnormal{G}$ ratios for the inter-core in Clump\,4 (color). The gray contours are the Stokes $I$ emission, and the contour levels are [25, 50, 100, 150, 200, 350, 500, 650] $\times$ $\sigma$ (1$\sigma$ = 48 $\mu$Jy\,beam$^{-1}$). For all panels, the black solid contours indicate the cores, the green solid contours define the inter-cores. The green stars within the cores denote their positions of mass centroids. The green arrows indicate the projection directions of the ratio of the two forces.} 
\label{fbg}    
\end{figure*}

Inspired by \cite{zhao2024}, we quantitatively compare the magnetic field tension force $F_\textnormal{B}$ and the gravitational force $F_\textnormal{G}$ between the cores (inter-core) as shown in \autoref{fbg} for Clump\,1 and Clump\,4, respectively. By using the KTH method \citep{koch2012a}, the magnetic field tension force is described as $F_{\textnormal{B}}=B^{2}/(4\pi R)=\left(\sqrt{\Sigma_{\textnormal{B}}(\rho\nabla \Psi)4\pi R}\right)^2/(4\pi R)=\Sigma_{\textnormal{B}}\rho\nabla \Psi$, in which $\Sigma_\text{B}$ is defined as $\Sigma_{\textnormal{B}}=\left(F_\textnormal{B}/|F_\textnormal{G}| \right)_{\textnormal{local}}=\sin\phi_\textnormal{LG}^\textnormal{NG}/\sin(90^{\circ}-\phi_\textnormal{B}^\textnormal{NG})$ \citep{koch2012a,koch2012b}, and the gravitational force is $F_\textnormal{G}=\rho\nabla \Psi$, if the hydrostatic dust pressure force is negligible. Here, $\rho=\mu{_{\textnormal{H}_{2}}}m{_{\textnormal{H}}}n{_{\textnormal{H}_{2}}}$ is the gas density, and $\nabla \Psi$ is the gradient of gravitational potential. The direction of $F_\textnormal{G}$ is calculated from $\nabla \Psi$. The direction of $F_\textnormal{B}$ is determined by first tracing the magnetic field lines based on the rotated polarization vectors, followed by a 90 degree rotation toward the concave side of the curved field lines \citep{koch2012a,zhao2024}. In general, the greater the curvature, the stronger the magnetic tension force. If the vectors are mostly aligned, it indicates little to no curvature, and the direction of the magnetic tension force becomes undefined (the pixels with undefined direction were masked out). Finally, we project the magnitude and direction of the forces within the inter-core along the direction toward the mass centroids of the cores. For each inter-core between a pair of cores, the force vectors are projected toward one core at a time. The negative values of $F_\textnormal{B}/F_\textnormal{G}$ indicate that the two forces act in opposite directions, whereas the positive values indicate that they act in the same direction.

The left and right panels of \autoref{fbg} show the same inter-cores projected onto the centroid mass of different cores. \autoref{fbg}\,(a) and (c) show that the $F_\textnormal{B}/F_\textnormal{G}$ ratios for the inter-cores toward Clump\,1-A and Clump\,4-A mostly fall within the range of $-$1 to 1, indicating that $F_\textnormal{B}$ is generally comparable to or weaker than $F_\textnormal{G}$ toward these two cores. On the other hand, \autoref{fbg}\,(b) and (d) show that more pixels in the inter-cores have $F_\textnormal{B}/F_\textnormal{G}$ in the range of $-$2 to $-$1 toward Clump\,1-B and Clump\,4-B, suggesting a stronger $F_\textnormal{B}$ acting in the opposition of $F_\textnormal{G}$. Consequently, the gas is more likely to infall toward the centers of Clump\,1-A and Clump\,4-A. We note that Clump\,1-A and Clump\,4-A are more massive than Clump\,1-B and Clump\,4-B. Thus, in the inter-cores, the magnetic field may provide some support against gravity, but it is insufficient to prevent gas from infalling toward the cores, particularly towards the more massive cores.

\subsection{Equilibrium States and Star Formation Activities of the Cores}

All the cores in Clump\,1 and Clump\,4 have shown $\alpha_{\textnormal{k}+\textnormal{B}}\lesssim1$, which suggest that these cores are gravitationally bound. Clump\,1-A and Clump\,1-B have shown $\alpha_{\textnormal{k}+\textnormal{B}}\sim1$, which means that these two cores are approaching a nearly quasi-equilibrium state \citep{liujh2024}, suggesting that the gravitation collapse is likely balanced by magnetic field and other supporting forces. Clump\,4 shows the relatively low values of $\alpha_{\textnormal{k}+\textnormal{B}}\sim0.3$, suggesting a sub-virial state where the gravitational force likely dominates gas dynamics in these cores. Clump\,1-A, Clump\,1-B, and Clump\,4-A show $\lambda>1$ and $\mathcal{M}_{\textnormal{A}}>1$, suggesting that both gravity and turbulence could shape the physical properties of these cores. For the diffuse Clump\,1-tail, both $\lambda$ and $\mathcal{M}_{\textnormal{A}}$ are smaller than 1. This could indicate that the magnetic field is dominant over both turbulence and gravity, probably preventing the gas from collapsing to form cores.

Both H$_{2}$O masers and outflows are clearly detected toward Clump\,1-A, Clump\,1-B, Clump\,4-A, Clump\,4-B and Clump\,4-C \citep{lu2019ApJS,lu2019,luxing2021}, along with the derived $\alpha_{\textnormal{k}+\textnormal{B}}<1$ (sub-viral), indicating ongoing star formation activities in these cores. There is no H$_{2}$O maser and outflow identified in Clump\,4-D \citep{lu2019ApJS,lu2019,luxing2021}. The sub-virial state and the predominantly perpendicular alignment of $\textnormal{AM}_{B}^{LG}$ suggest there is no clear sign indicating star formation activities in Clump\,4-D. Clump\,4-E has shown signatures of star formation, because it is in a sub-virial state with outflow detection, although there is no maser detection toward this core \citep{lu2019,lu2019ApJS,luxing2021}. Oppositely, Clump\,4-F was only detected in H$_{2}$O maser without outflow detection. Moreover, the $\textnormal{AM}_{B}^{LG}$ shows the  predominantly parallel alignment. These suggest Clump\,4-F might reach the pre-collapse or the early collapse phase, yet star formation has not begun.

\section{Conclusions}
We present the results of ALMA dust polarization observations to show the magnetic field structures of Clump\,1 and Clump\,4 in the 20\,km\,s$^{-1}$ cloud. In addition, we also study the multi-scale magnetic fields with the JCMT polarization data \citep{yang2025}. The results are summarized as follows.

\begin{enumerate}
\item We have divided both Clump\,1 and Clump\,4 into multiple cores using \texttt{astrodendro}, and Clump\,1-tail is an additional diffuse substructure enclosed at 5$\sigma$ level in Clump\,1. The magnetic field strengths derived using the ADF method range from 0.3$-$3.1\,mG. The normalized mass-to-flux ratios $\lambda$ for the cores in Clump\,1 and Clump\,4 (except for Clump\,1-tail) are $\lambda>1$, suggesting that these cores are mostly magnetically supercritical, where gravity dominates over magnetic support, while Clump\,1-tail is in the subcritical state, in which magnetic support overcome the gravity. All the cores in Clump\,1 and Clump\,4 have shown $\alpha_{\textnormal{k}+\textnormal{B}}\lesssim1$, which suggest that these cores are likely gravitationally bound. All the cores in Clump\,1 and Clump\,4 are super-Alvénic ($\mathcal{M}_{\textnormal{A}}\gtrsim1$), while Clump\,1-tail sub-Alvénic ($\mathcal{M}_{\textnormal{A}}\sim0.7$), suggesting that turbulence plays a more important role than magnetic field in the dense cores and a less dominant role in the diffuse region.

\item We study the angular difference between multi-scale magnetic fields probed by JCMT and ALMA. The results suggest that the cloud-scale (0.6 pc) and core-scale (5000 au) magnetic fields show angular deviations, with the primary difference arising from the dense cores. At the cloud-scale, the magnetic field predominates, while gravity becomes the dominant force at the core-scale. The magnetic field orientations are largely consistent between the core-scale and condensation-scale (2000 au), indicating that both are predominantly governed by gravity.

\item The relative orientation between the magnetic field and local gravity ($\phi_\textnormal{B}^\textnormal{LG}$) suggests gravity plays an important role in shaping the dynamics in the diffuse Clump\,1-tail. The magnetic field of the dense cores is likely affected by both gravity and star formation activities.

\item The ratio of the magnetic field tension force $F_\textnormal{B}$ to the gravitational force $F_\textnormal{G}$ for inter-cores in Clump\,1 and Clump\,4 suggests that the magnetic field may provide some support against gravity, but it is insufficient to prevent gas from infalling toward the cores.

\end{enumerate}

\section*{Acknowledgments}
We thank the referee for the thorough review and for providing the helpful and detailed comments. This work benefits from prior interactions with authors of \citet{zhao2024}. They established the accuracy of the acceleration vectors and the magnetic force orientation estimation using Magnetohydrodynamics (MHD) simulation data. Y.H.L. thanks all members of the Star Formation group in Shanghai Astronomical Observatory for their help and support. This work has been supported by the Natural Science Foundation of Shanghai (No.\ 23ZR1482100), the Strategic Priority Research Program of the Chinese Academy of Sciences (CAS) Grant No.\ XDB0800300, the National Natural Science Foundation of China (NSFC) through grant Nos.\ 12273090 and 12322305, the National Key R\&D Program of China (No.\ 2022YFA1603101), and the CAS ``Light of West China'' Program No.\ xbzg-zdsys-202212. PSL acknowledges the support by the National Key R\&D Program of China (No. 2022YFA1603100) and National Natural Science Foundation of China (NSFC) through grant No.\ 1241101426. M.Z.Y. and S.P.L. acknowledge support from the National Science and Technology Council (NSTC) of Taiwan under grants 112-2112-M-007-011, 113-2112-M-007- 004, and 114-2112-M-007-001. This paper makes use of the following ALMA data: ADS/JAO.ALMA\#2021.1.00286.S. ALMA is a partnership of ESO (representing its member states), NSF (USA), and NINS (Japan), together with NRC (Canada), MOST and ASIAA (Taiwan), and KASI (Republic of Korea), in cooperation with the Republic of Chile. The Joint ALMA Observatory is operated by ESO, AUI/NRAO, and NAOJ. \\
\facilities{ALMA}
\software{astropy \citep{astropy2013}, APLpy \citep{robitaille2012}, CASA \citep{bean2022casa}, matplotlib \citep{hunter2007}, astrodendro \citep{rosolowsky2008}}

\bibliography{reference}
\bibliographystyle{aasjournalv7}

\appendix
\section{Dendrogram parameters}\label{apdx-astro}
To better analyze the physical and the magnetic field properties, we divided each clump into cores using the \texttt{astrodendro} Python package \citep{Robitaille2019,rosolowsky2008}. The cores were identified as branches, which represent the intermediate structures that can split into smaller branches or leaves in the hierarchical structures of the dendrogram. When applying \texttt{astrodendro}, the minimum value (\texttt{min$\_$value}) was set to 10\,$\sigma$, the minimum significance for structures (\texttt{min$\_$delta}) was set at 200\,$\sigma$, and the minimum number of pixels (\texttt{min$\_$npix}) that required for a substructure was set to 35 pixels, which is equivalent to the number of pixels inside one beam. The \texttt{min$\_$delta} was set sufficiently high to avoid leaves (with no further substructures in the hierarchy) from being classified as cores. This parameter setup for core identification was determined through trial and error to ensure consistency with the ALMA C-2 configuration data at the core-scale ($\sim$5000 au; see \autoref{sec:multi-scale}). Additionally, we manually enclosed the northeastern side of the Clump\,1 at 5\,$\sigma$ to study the morphology of the diffuse emission and named it as Clump\,1-tail as shown in \autoref{fig1-cont-B-vec}\,(b). The condensations were identified as the isolated compact structures within each core.

\section{Velocity dispersion and angle dispersion}\label{apdx-vel_ang}

\autoref{v-dispersion} shows the velocity dispersion for Clump\,1 and Clump\,4 obtained by performing Gaussian fittings. When choosing the molecular line to derive the velocity dispersion, we selected a line with a critical density that is more or less comparable to the density of the measured region. For instance, for the diffuse region such as the Clump\,1-tail, we used the NH$_{3}$ ($J$,$K$ = 3,3) line from Karl G.\ Jansky Very Large Array (JVLA) observations \citep{lu2017}, and for the relatively dense cores, we used the HN$^{13}$C\,($J$ = 4$-$3) line from the same ALMA observations of this work but excluding the QA-semipass data to derive the velocity dispersion. The velocity dispersion was calculated by first performing a Gaussian fit to each pixel individually within all the cores and Clump\,1-tail, then taking the average of the standard deviations across all pixels within each core and Clump\,1-tail as the dispersion value. \autoref{clump-angle} shows the polarization position angle distributions for the entire Clump\,1 and Clump\,4. \autoref{sigmat} shows the polarization position angle distributions for the cores within each clump. The polarization angle dispersions are obtain by performing the Gaussian fittings of the angle distributions.

\renewcommand{\thefigure}{B\arabic{figure}}
\renewcommand{\theHfigure}{B\arabic{figure}}
\setcounter{figure}{0}

\begin{figure*}[ht]
\gridline{\hspace{-2\baselineskip}
          \fig{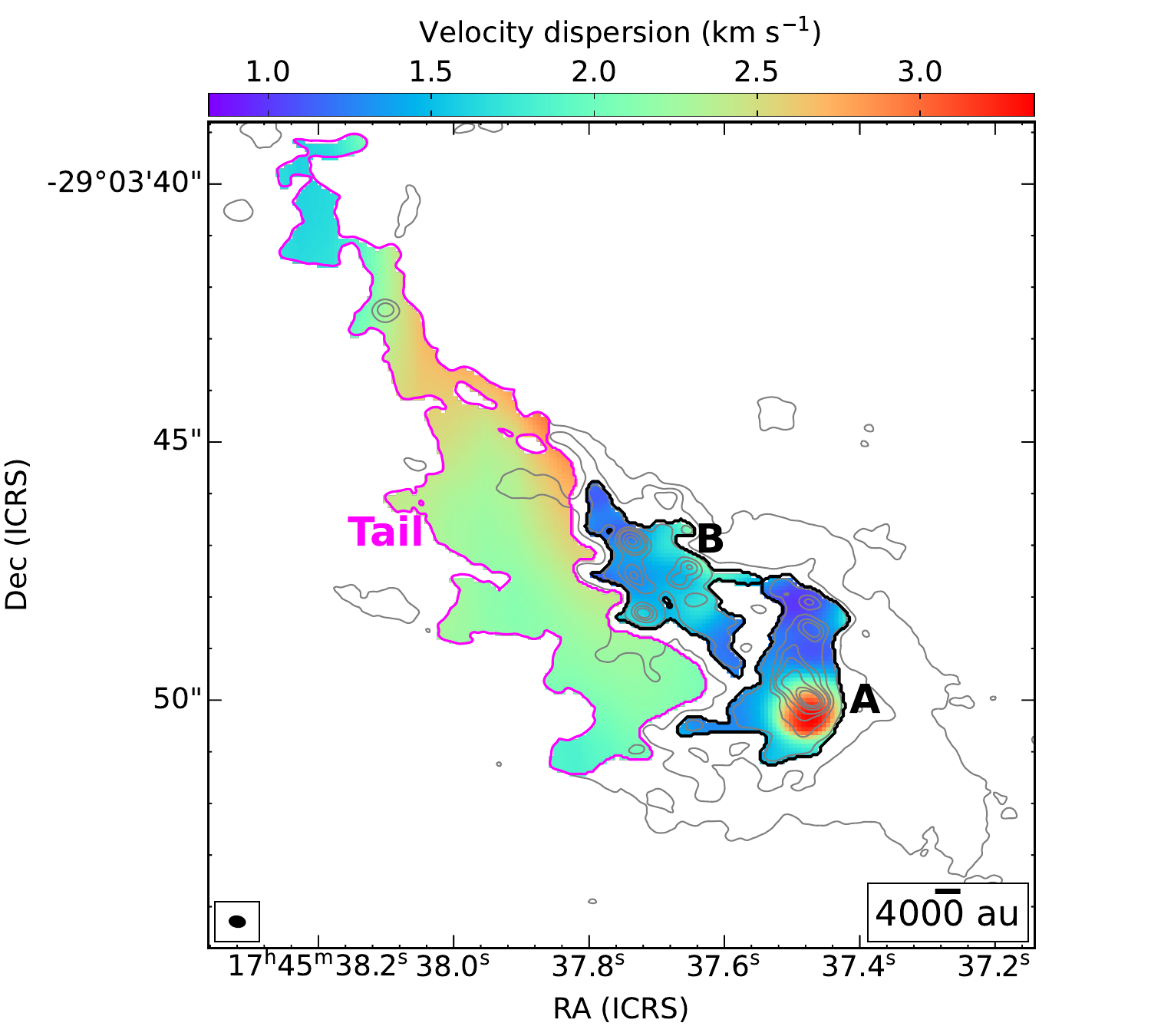}{0.5\textwidth}{}
          {}\hspace{-2\baselineskip}
           \fig{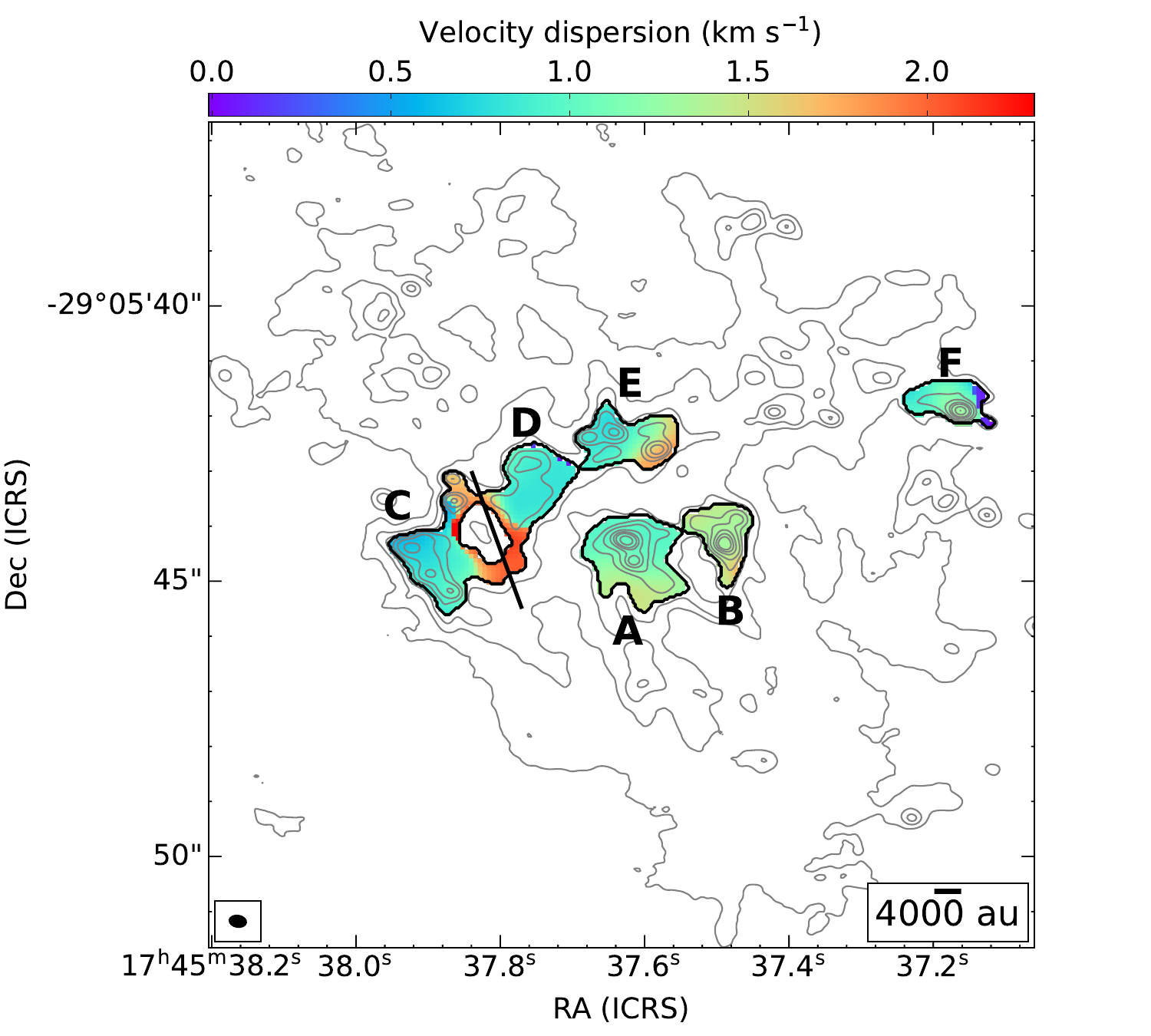}{0.5\textwidth}{}
          }

\vspace{-2\baselineskip}
\caption{Left panel: The color shows the velocity dispersions for Clump\,1-A and Clump\,1-B estimated using the HN$^{13}$C\,($J$ = 4$-$3) emission and for Clump\,1-tail estimated using the NH$_{3}$ emission. The grey contours are the Stokes $I$ and the contour levels in both panels are identical to \autoref{fig1-cont-B-vec}\,(b). Right panel: The color shows the velocity dispersions for Clump\,4 estimated using the HC$_{3}$N emission. The grey contours are the Stokes $I$ and the contour levels are identical to \autoref{fig1-cont-B-vec}\,(c). The cores enclosed by black solid contours in both panels are identified by \texttt{astrodendro}, and the magenta contour in left panel denotes Clump\,1-tail.} 
\label{v-dispersion}       
\end{figure*}

\begin{figure*}[ht]
\gridline{\hspace{-2\baselineskip}
          \fig{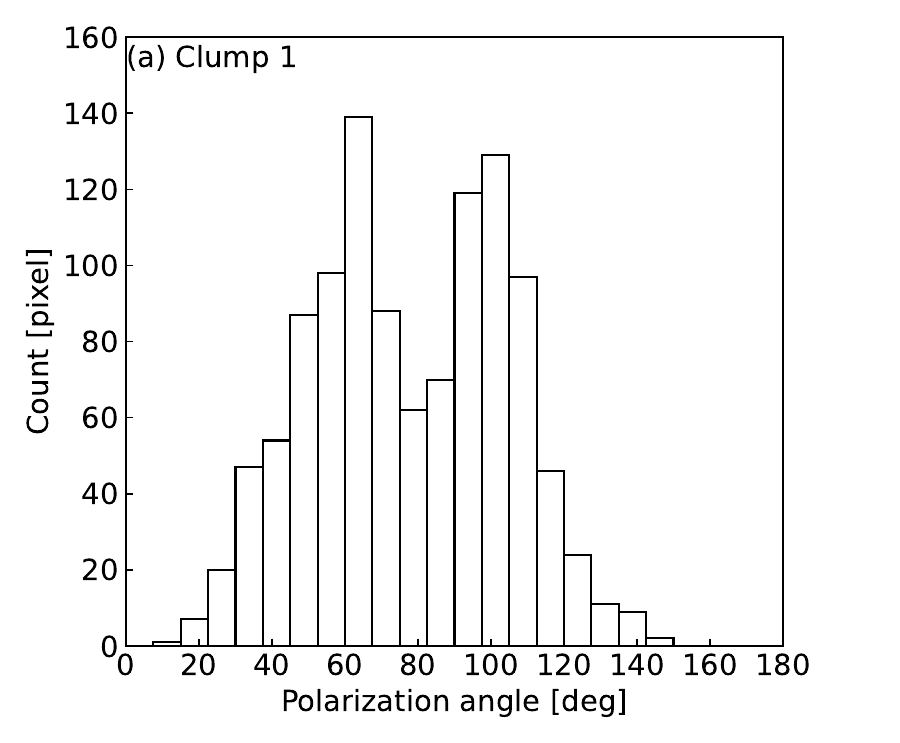}{0.32\textwidth}{}{}\hspace{-5\baselineskip}
          \fig{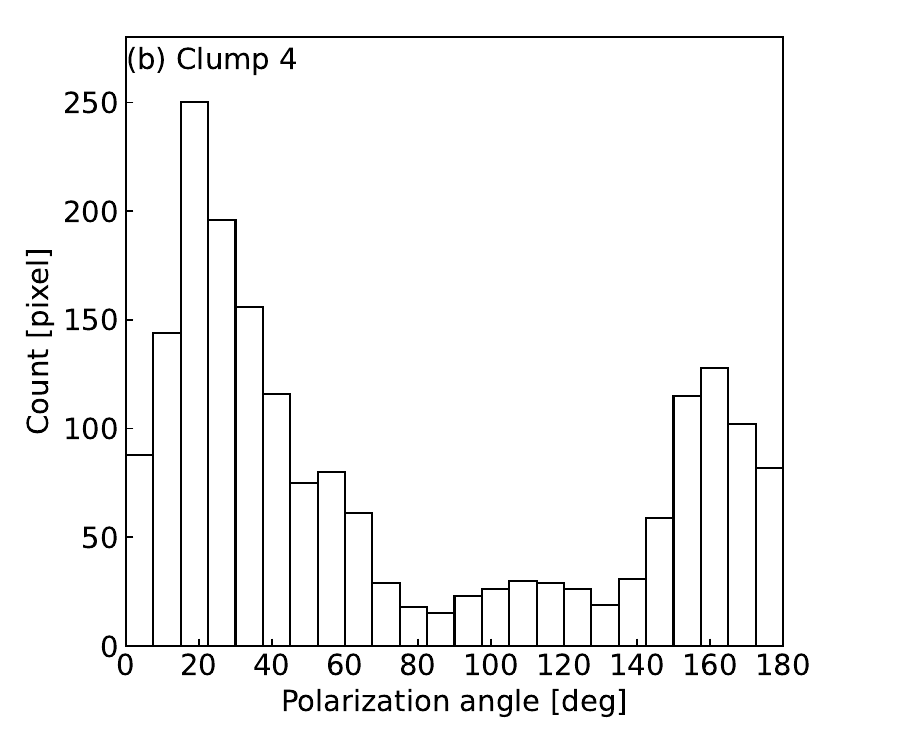}{0.32\textwidth}{}{}
          }
\vspace{-2\baselineskip}
\caption{Histograms of the polarization position angles for the Clump\,1 and Clump\,4. The polarization position angles are measured from north toward east.} 
\label{clump-angle}
\end{figure*}

\begin{figure*}[ht]
\gridline{\hspace{-3\baselineskip}
          \fig{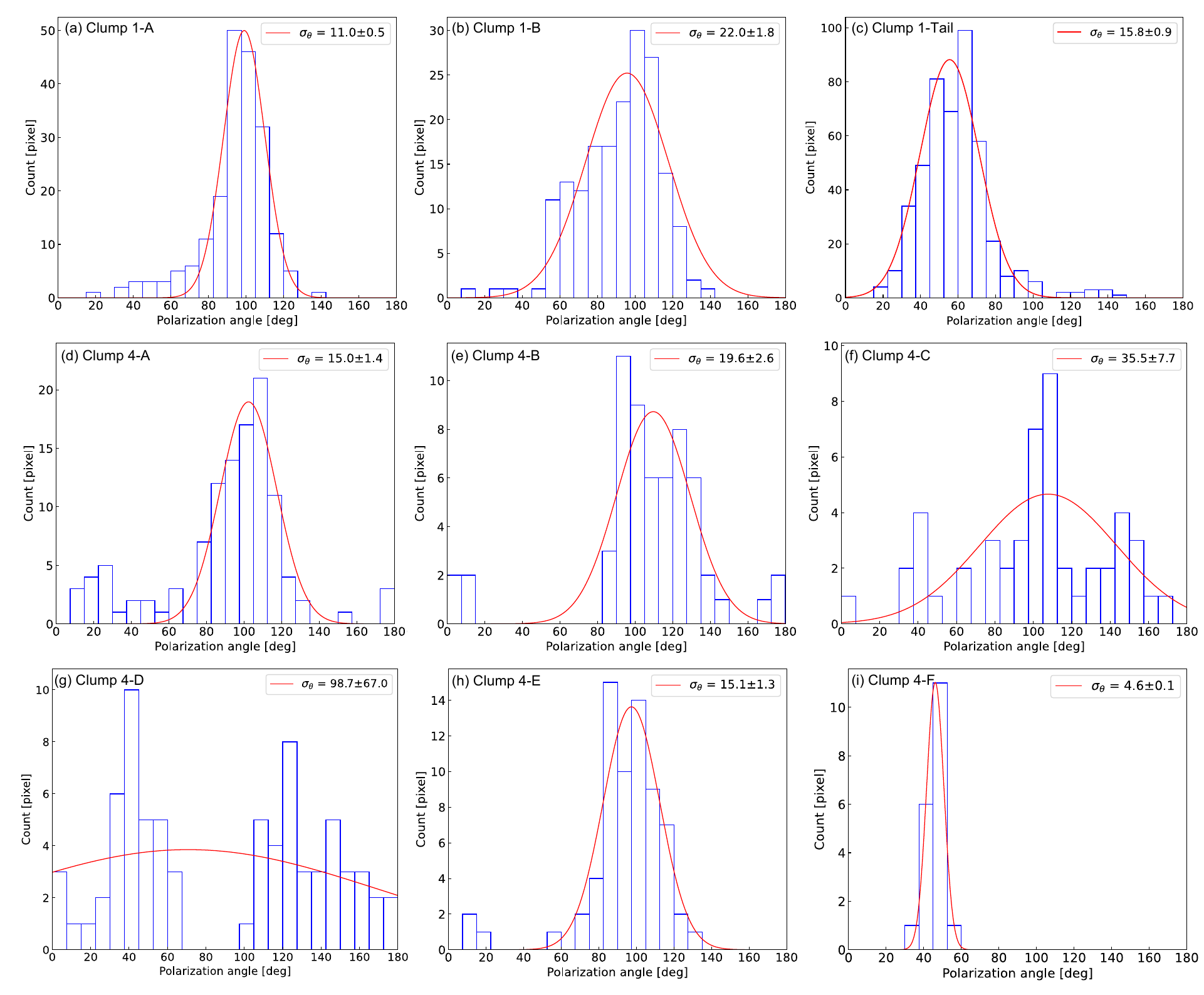}{1\textwidth}{}{}\hspace{-3\baselineskip}}
\vspace{-2\baselineskip}
\caption{Histograms of the polarization position angles for the cores of Clump\,1 and Clump\,4. The polarization position angles are measured from north toward east. The red curve denotes the Gaussian fitting line in each panel.} 
\label{sigmat}
\end{figure*}

\section{Magnetic field strength estimation with ADF method}\label{apdx-B-field}

\renewcommand{\theequation}{C\arabic{equation}}
\renewcommand{\theHequation}{C\arabic{figure}}
\setcounter{equation}{0}

\renewcommand{\thefigure}{C\arabic{figure}}
\renewcommand{\theHfigure}{C\arabic{figure}}
\setcounter{figure}{0}

\begin{figure*}[ht]
\gridline{\hspace{-3\baselineskip}
          \fig{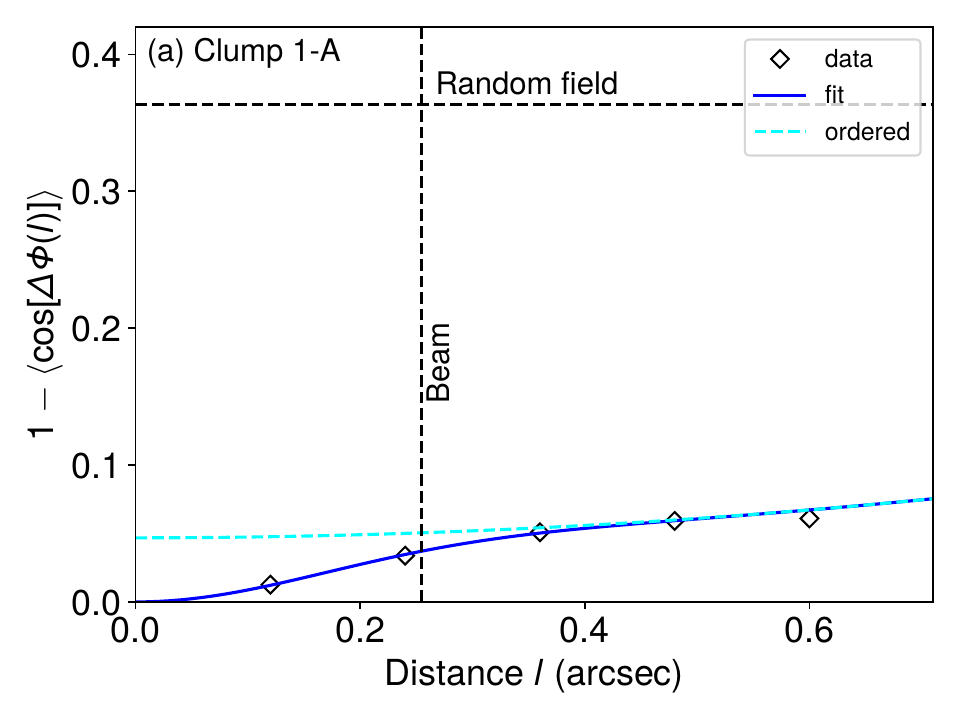}{0.32\textwidth}{}
          {}\hspace{-3\baselineskip}
          \fig{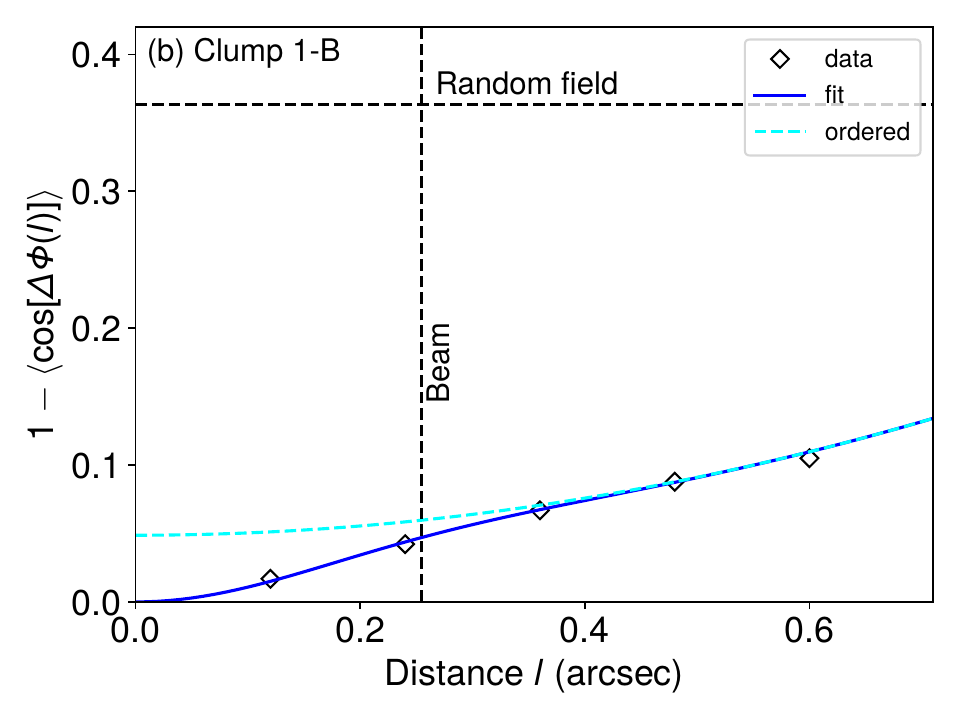}{0.32\textwidth}{}
          {}\hspace{-3\baselineskip}
          \fig{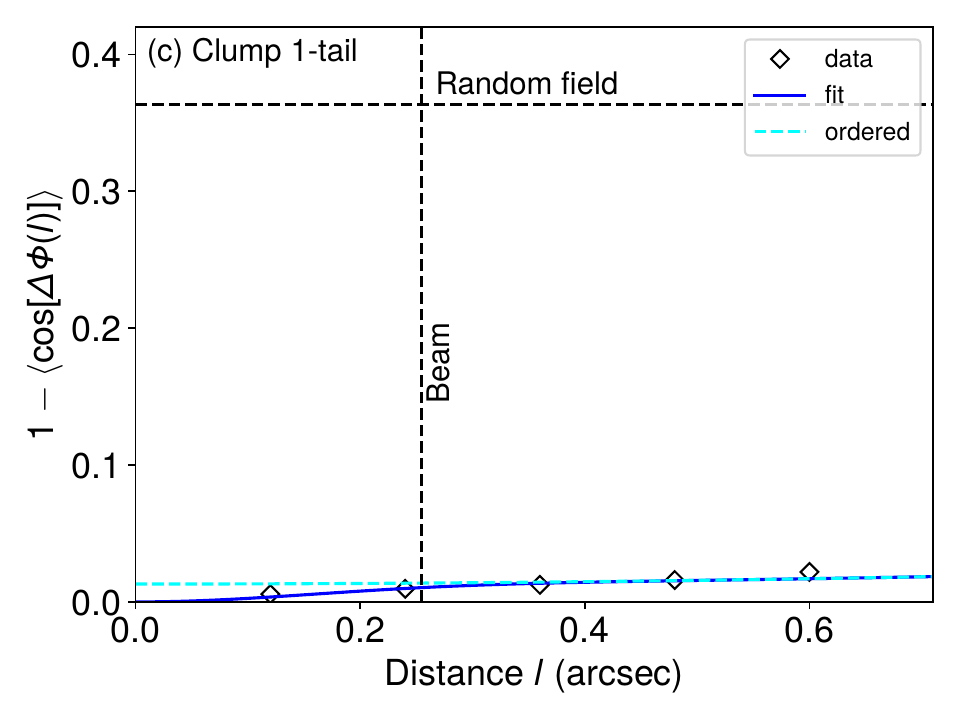}{0.3\textwidth}{}
          {}\hspace{-3\baselineskip}}
\vspace{-2\baselineskip}
\gridline{\hspace{-4\baselineskip}
          \fig{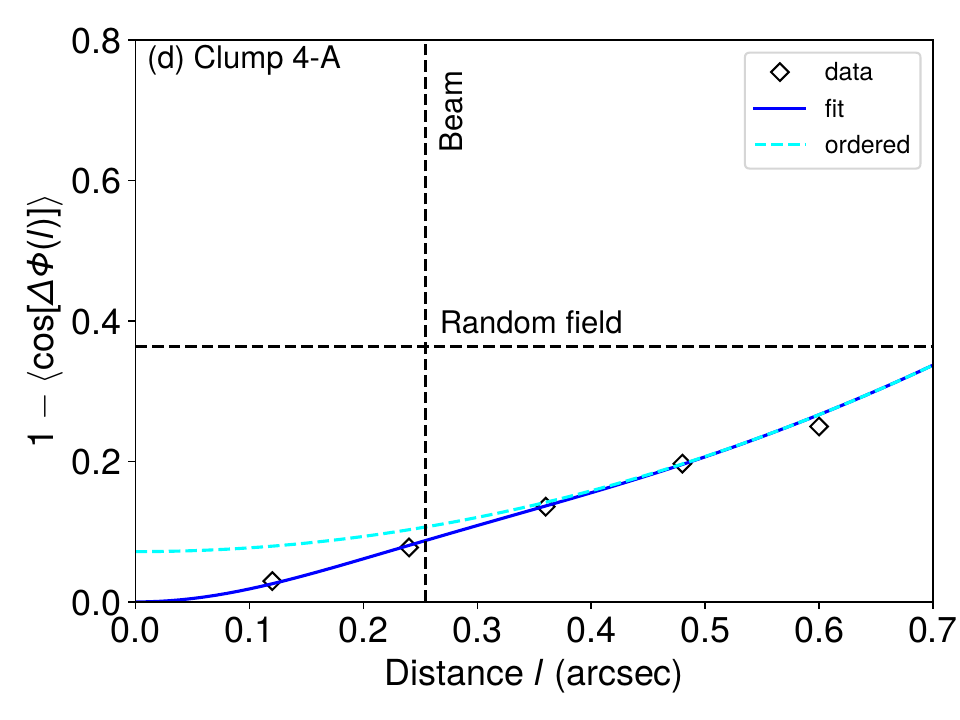}{0.3\textwidth}{}
          {}\hspace{-3\baselineskip}
          \fig{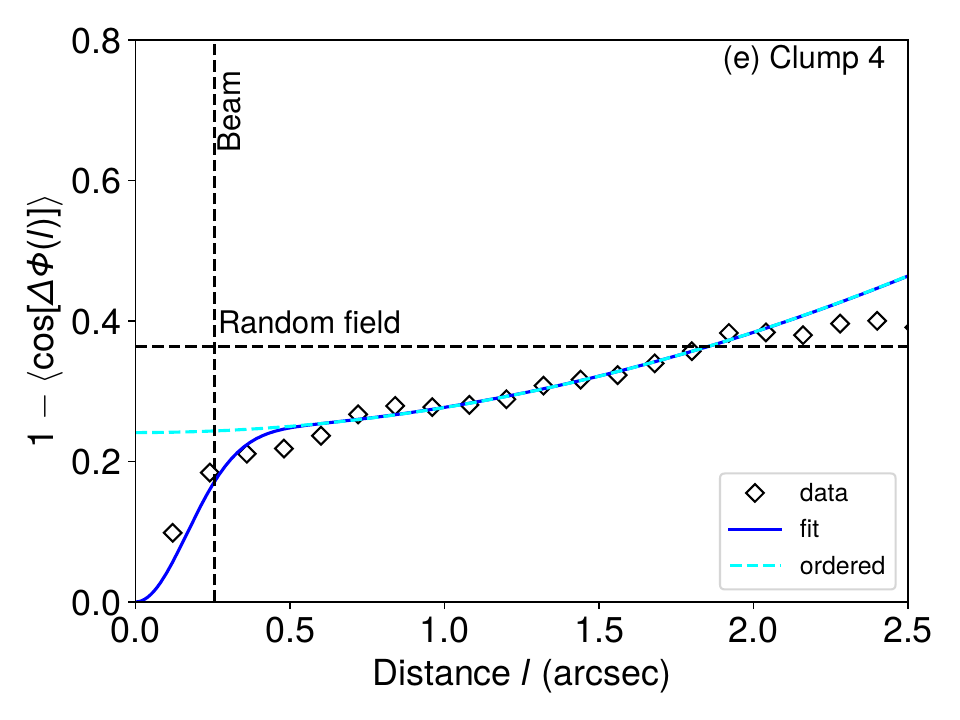}{0.3\textwidth}{}
          {}\hspace{-3\baselineskip}
          }
\vspace{-2\baselineskip}
\caption{The angular dispersion function (ADF) $1-\langle\textnormal{cos}[\Delta\Phi(l)]\rangle$ (in diamonds) is plotted as a function of the distance ($l$). The blue line shows the correlated turbulent component of the best fit. The cyan labels within the plots indicate the ordered component due to the synthesized beam. The vertical black dashed line shows the synthesized beam size. The horizontal dashed line shows the ADF value corresponding to a random field \citep{liujunhao2021}.} 
\label{fig-adf}
\end{figure*}

The ADF method improves upon the Davis–Chandrasekhar–Fermi (DCF) method in the angular dispersion estimation by taking into account of the ordered field contribution, observational effects, and line-of-sight signal averaging \citep{hildebrand2009,houde2009,houde2016}. The ADF shown in \autoref{fig-adf} is given by \cite{houde2016}: 
\begin{equation}
\begin{multlined}
1-\langle\textnormal{cos}[\Delta\Phi(l)]\rangle\simeq a_{2}l^{2}+ \frac{\langle B_{0}^{2} \rangle}{\langle B_{\textnormal{t}}^{2} \rangle} C \times \\ \Bigl\{
\frac{1}{C_{1}}[1-e^{-l^{2}/2(\delta^{2}+2W_{1}^{2})}] + \\
\frac{1}{C_{2}}[1-e^{-l^{2}/2(\delta^{2}+2W_{2}^{2})}]- \\
\frac{1}{C_{12}}[1-e^{-l^{2}/2(\delta^{2}+W_{1}^{2}+W_{2}^{2})}]
\Bigr\},
\end{multlined}
\end{equation}
where $\Delta\Phi(l)$ is the angular difference of two magnetic field vectors at a separation of distance $l$, $a_{2}l^{2}$ is the second-order term of the Taylor expansion for the ordered field, $\delta$ is the turbulent correlation length, $B_{\textnormal{t}}$ is the turbulent magnetic field, $B_{0}$ is the ordered magnetic field, $W_{1}$ is the geometric mean of FWHM of the synthesized beam size divided by $\sqrt{8\, \textnormal{ln}\,2}$, and $W_{2}$ accounts for the large-scale filter effect obtained by using the maximum recoverable scale divided by $\sqrt{8\, \textnormal{ln}\,2}$. $C$ is the coefficient, which is given by \citet{houde2016}:
\begin{equation}
C_{1}=\frac{(\delta^2+2W_{1}^{2})}{\sqrt{2\pi}\delta^{3}},
\end{equation}

\begin{equation}
C_{2}=\frac{(\delta^2+2W_{2}^{2})}{\sqrt{2\pi}\delta^{3}},
\end{equation}

\begin{equation}
C_{12}=\frac{(\delta^2+W_{1}^{2}+W_{2}^{2})}{\sqrt{2\pi}\delta^{3}},
\end{equation}
and
\begin{equation}
C=(\frac{1}{C_{1}}+\frac{1}{C_{2}}-\frac{2}{C_{12}})^{-1}.
\end{equation}

The turbulent-to-total field strength ratio $ \langle B^{2}_{t} \rangle/\langle B^{2} \rangle$ is given by \citet{liujunhao2021}:
\begin{equation}
\langle B^{2}_{\textnormal{t}}\rangle/\langle B^{2} \rangle = \frac{\langle B^{2}_{\textnormal{t}}\rangle/\langle B_{0}^{2}\rangle}{1+(\langle B^{2}_{\textnormal{t}}\rangle/\langle B^{2}_{0} \rangle)}.
\end{equation}

\section{VLA C-band continuum image}\label{apdx-Hii}

\renewcommand{\thefigure}{D\arabic{figure}}
\renewcommand{\theHfigure}{D\arabic{figure}}
\setcounter{figure}{0}

Previous study has shown that the magnetic field structures in the CMZ are likely affected by its physical environments such as H\textsc{ii} regions \citep{lu2024}. Key findings include the identification of the cometary magnetic field in Sgr C possibly due to the cloud interaction with an expanding H\textsc{ii} region, and the curved magnetic field in the Dust Ridge possibly tracing converging gas flows \citep{lu2024}. 

\begin{figure*}[ht]
\gridline{\hspace{-2\baselineskip}
          \fig{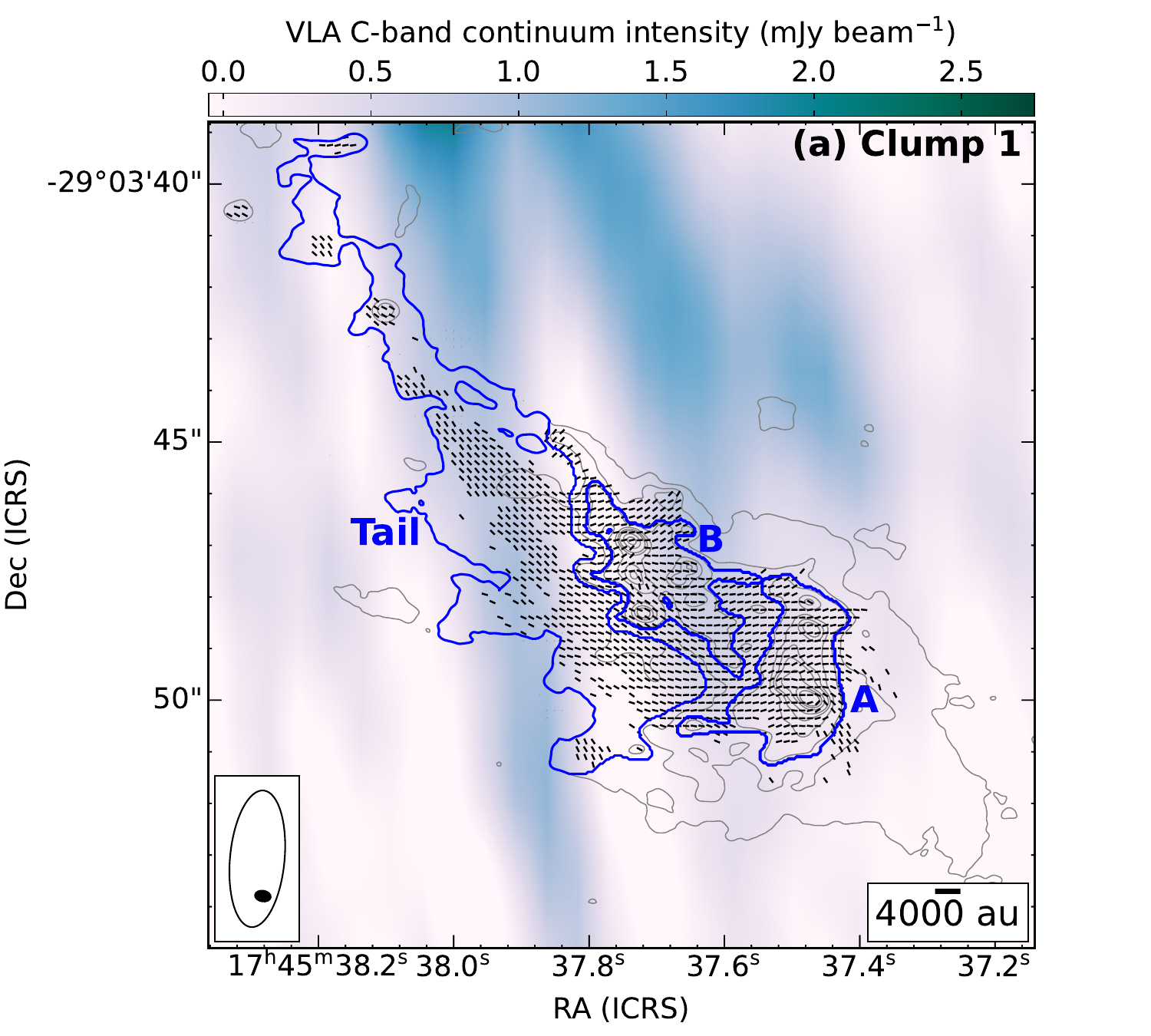}{0.55\textwidth}{}
          \fig{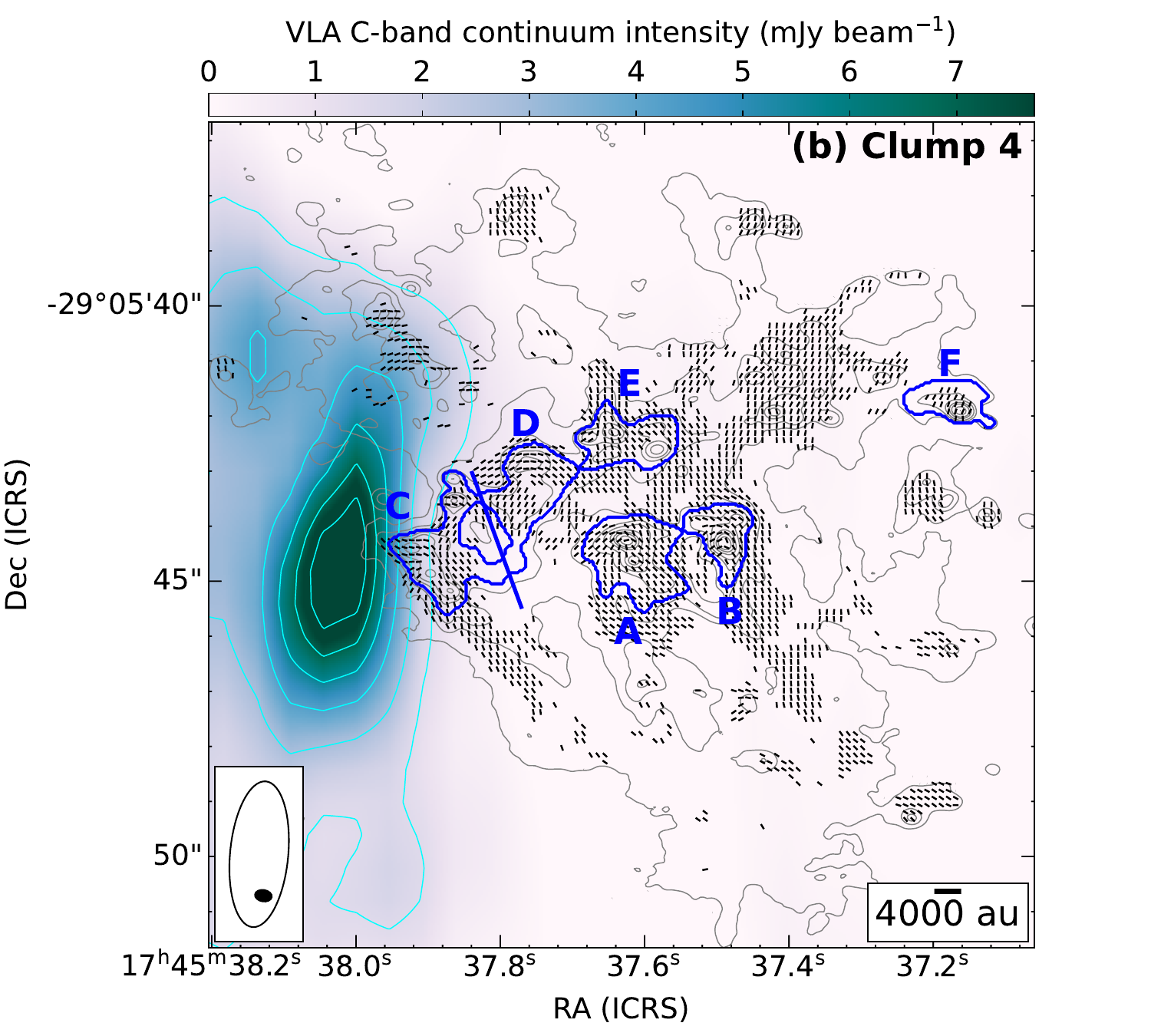}{0.55\textwidth}{}     
          }
\vspace{-1\baselineskip}
\caption{VLA C-band continuum emission \citep{lu2019ApJS} in color and cyan contours overlaid with the ALMA 870\,$\mu$m Stokes $I$ emission (this work) in gray contours and the inferred magnetic field orientations (this work) in black line segments for (a) Clump\,1 and (b) Clump\,4, respectively. The gray contour levels and black vectors in panels (a) and (b) are identical to those in the \autoref{fig1-cont-B-vec}. The cyan contour levels in panel (b) are [5, 10, 15, 20, 25, 30] $\times$ $\sigma$ (1$\sigma$ = 0.28 Jy\,beam$^{-1}$). The blue contours in panels (a) and (b) denote the cores identified by \texttt{astrodendro} and Clump\,1-tail. The synthesized beams for VLA and ALMA are denoted by the open ellipse and the black dot at the bottom left corner in each panel.}
\label{cband}
\end{figure*}

\cite{lu2015} and \cite{lu2017} have studied an H\textsc{ii} region in Clump\,4 as shown in \autoref{cband}\,(b). The magnetic field in Clump\,4-C across the 3$\sigma$ contour of the H\textsc{ii} region shows an orthogonal orientation. The vectors in Clump\,4-C on the western side of the 3$\sigma$ contour show a curvature structure along the contour line.

There is no detection of H\textsc{ii} regions toward Clump\,1 (\autoref{cband}\,(a)). However, several filaments (narrowly elongated structure shown in blueish color) are observed toward Clump\,1. The magnetic field orientation in Clump\,1-tail deviates from long axis of the filaments, which is partially consistent with the findings of \citet{pare2025}, who reported a perpendicular alignment between the magnetic field and the filament.

\end{CJK}
\end{document}